\documentclass[aip, reprint]{revtex4-1}

\usepackage{siunitx}
\usepackage{graphicx}
\usepackage{color}
\usepackage{amsmath}
\usepackage{amssymb}
\usepackage{microtype}
\usepackage[normalem]{ulem}

\begin{document}

\title{Liquid exfoliation of multilayer graphene in sheared solvents: a molecular dynamics investigation} 

\author{Simon Gravelle}
\affiliation{School of Engineering and Material Science, Queen Mary University of London, United Kingdom}
\author{Catherine Kamal} 
\affiliation{School of Engineering and Material Science, Queen Mary University of London, United Kingdom}
\author{Lorenzo Botto}\email{l.botto@tudelft.nl}
\affiliation{School of Engineering and Material Science, Queen Mary University of London, United Kingdom}
\affiliation{Process and Energy Department, 3ME Faculty of Mechanical, Maritime and Materials Engineering, TU Delft, Delft, The Netherlands}

\begin{abstract}
Liquid-phase exfoliation, the use of a sheared liquid to delaminate graphite into few-layer graphene, is a promising technique for the large-scale production of graphene. But the micro and nanoscale fluid-structure processes controlling the exfoliation are not fully understood. Here we perform non-equilibrium molecular dynamics simulations of a defect-free graphite nanoplatelet suspended in a shear flow and measure the critical shear rate $\dot \gamma_c$ needed for the exfoliation to occur. We compare $\dot \gamma_c$ for different solvents including water and NMP, and nanoplatelets of different lengths. Using a theoretical model based on a balance between the work done by viscous shearing forces and the change in interfacial energies upon layer sliding, we are able to predict the critical shear rates $\dot \gamma_c$ measured in simulations. We find that an accurate prediction of the exfoliation of short graphite nanoplatelets is possible only if both hydrodynamic slip and the fluid forces on the graphene edges are considered, and if an accurate value of the solid-liquid surface energy is used. The commonly used``geometric-mean" approximation for the solid-liquid energy leads to grossly incorrect predictions.
\end{abstract}

\maketitle

\section{Introduction}

\noindent Two-dimensional materials are made of a single layer of atoms and show physical properties not accessible with bulk materials \cite{Mas-Balleste2011, Mounet2018}. In particular, charge and heat transport confined to a plane display unusual behaviour \cite{Butler2013}. Among the family of two-dimensional materials, graphene is considered the thinnest and strongest material ever measured \cite{Geim2010}. Graphene possesses outstanding electrical, transport and thermal properties, and is an appealing candidate for numerous applications in fields such as electronics \cite{Avouris2012}, energy generation and storage \cite{Brownson2011}, or in biomedicine \cite{Chung2013}. However, the fabrication of single or few-layer graphene at the industrial scale remains a challenge.

\vspace{0.1cm}

Liquid-phase exfoliation is a promising technique for the large-scale production of graphene \cite{Yi2015}. It consists in dispersing microparticles of graphite in a liquid and forcing the separation of the particles into fewer layer graphene by using a large shear flow \cite{Hernandez2008, Coleman2013, Yi2014}. For rigid platelets, the exfoliation is expected to occur if the work of the hydrodynamic forces applied by the liquid on the layered particles is larger than the change in energy associated with the dissociation of the layers \cite{Chen2012, Paton2014}. The objective of the present article is to quantify this statement using molecular dynamics.

\vspace{0.1cm}

\begin{figure}
\includegraphics[width=\linewidth]{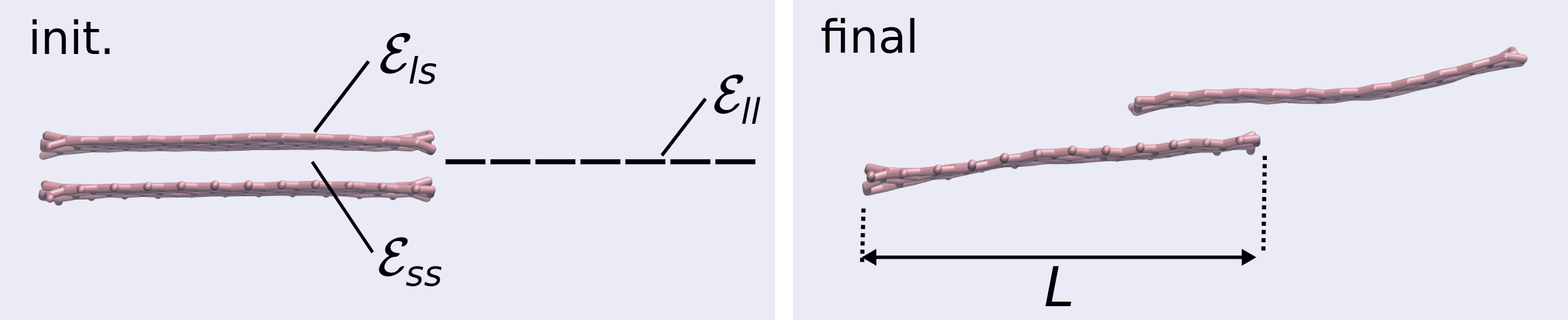}
\caption{Schematic of a bilayer nanoparticle of length $L$ in a liquid before exfoliation (init.), with the three different surface energy terms, and the same nanoparticle after exfoliation into two single-layer platelets (final).
}
\label{fig:FIGURE1}
\end{figure}

The change in energy associated with the separation of two layers in a liquid can be estimated following a model originally proposed by Chen et al. \cite{Chen2012} and later improved by Paton et al. \cite{Paton2014}. One considers a bilayer nanoplatelet of length $L$ and width $w$ immersed in a liquid. The total surface energy of the bilayer particle before exfoliation is 
\begin{equation}
\label{eq:EA}
E_\text{init} = - L w \left( {\cal E}_{ss} + {\cal E}_{\ell \ell} + 2 {\cal E}_{\ell s} \right),
\end{equation}
where ${\cal E}_{ss}$, ${\cal E}_{\ell \ell}$, and ${\cal E}_{\ell s}$, are the solid-solid, liquid-liquid, and liquid-solid surface energy densities respectively (Fig.\,\ref{fig:FIGURE1}). After exfoliation, the total surface energy of the separated particles is (Fig.\,\ref{fig:FIGURE1})
\begin{equation}
\label{eq:EC}
E_\text{final} = - 4 L w {\cal E}_{\ell s}.
\end{equation}
The total change in energy $\Delta E = E_\text{final} - E_\text{init}$ associated with the particle exfoliation is thus
\begin{equation}
\label{eq:DEfull}
\Delta E = L w \left( {\cal E}_{ss} + {\cal E}_{\ell \ell} - 2 {\cal E}_{\ell s} \right).
\end{equation}
Since ${\cal E}_{\ell s}$ is not known in general, the geometric mean approximation ${\cal E}_{\ell s} = \sqrt{{\cal E}_{\ell \ell} {\cal E}_{s s}}$ connecting the solid-liquid to the liquid-liquid and solid-solid surface energies is commonly used, resulting in
\begin{equation}
\label{eq:DE}
\Delta E = L w \left( \sqrt{{\cal E}_{\ell \ell}} - \sqrt{{\cal E}_{s s}} \right)^2.
\end{equation}

\vspace{0.1cm}

\begin{figure}
\includegraphics[width=\linewidth]{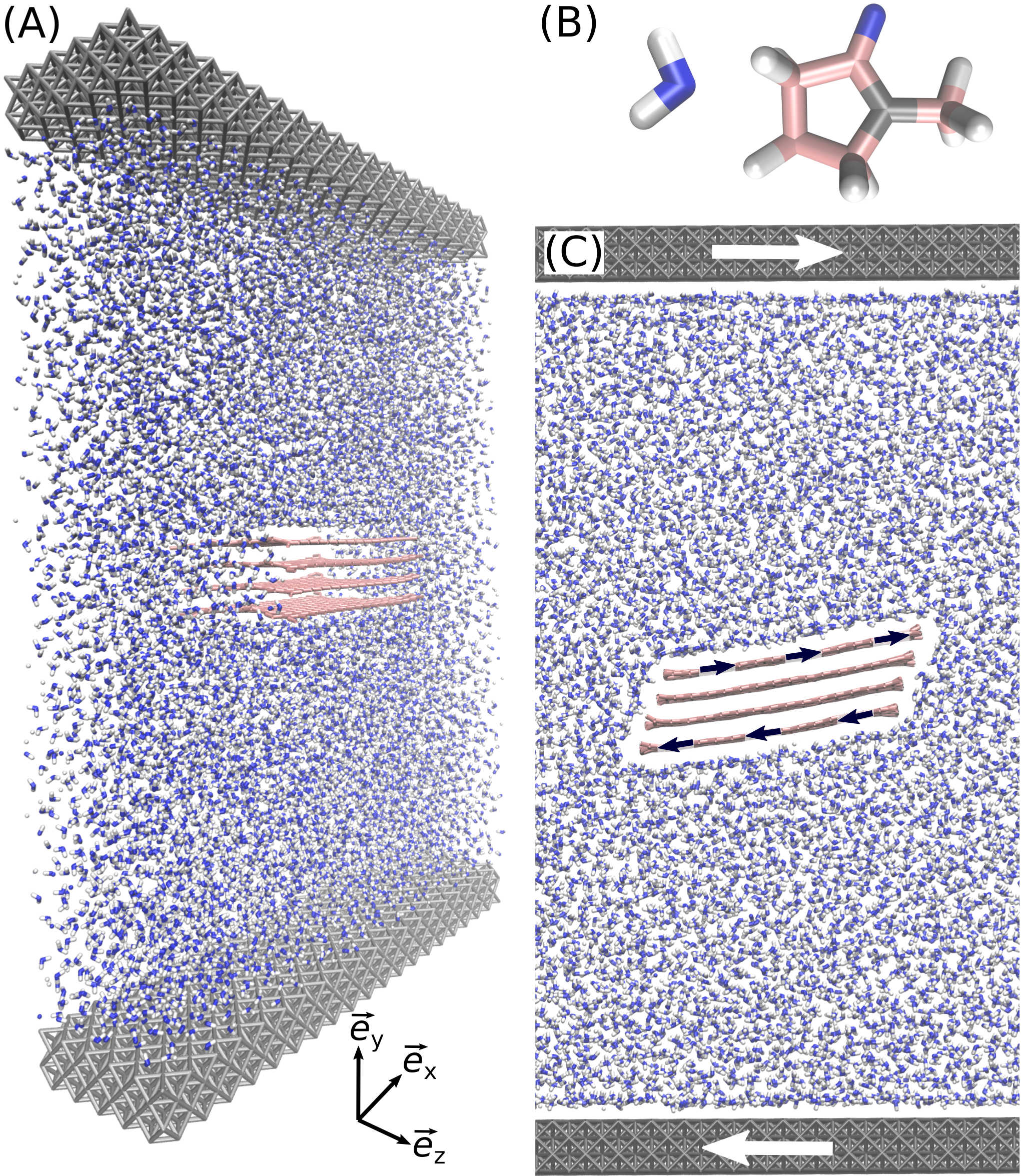}
\caption{
\textbf{(A-C)} Molecular dynamics snapshots of a four-layer graphite nanoparticle prior to exfoliation in water. The particle has size dimension $L$ in the ${\vec e}_x$ direction and height $H$ in the ${\vec e}_y$ direction \cite{VMD}. The third dimension in the ${\vec e}_z$ direction is $w$. Two moving walls impose a linear shear flow profile on the liquid. The black arrows indicate qualitatively the direction of the tangential hydrodynamic force applied to the top and bottom layers on the nanoparticle.
\textbf{(B)} Water (left) and NMP (right) molecules.
}
\label{fig:FIGURE2}
\end{figure}

Exfoliation is expected if the work done by the tangential hydrodynamics force ${\cal W}_\text{hyd}$ applied by the shearing liquid on the particle is larger than $\Delta E$. Assuming the no-slip boundary condition and ignoring contributions from the edges of the platelet, the tangential hydrodynamic force driving the relative sliding of the top and bottom layers is $F_\text{hyd} \approx \eta \dot \gamma w L$, where $\eta$ is the fluid viscosity, $\dot \gamma$ the shear rate applied to the fluid \cite{Singh2014}. The total work required for the hydrodynamic force to separate one layer from the other in a sliding deformation can be estimated as 
\begin{equation}
\label{eq:wf}
{\cal W}_\text{hyd} = F_\text{hyd} L \approx \eta \dot \gamma w L^2.
\end{equation}
By equating Eq.\,\eqref{eq:DE} and Eq.\,\eqref{eq:wf}, the following expression for the critical shear rate value $\dot \gamma_c$ above which exfoliation is expected is obtained
\begin{equation}
\label{eq:gammaStar}
\dot \gamma_c \approx \dfrac{1}{\eta L} \left( \sqrt{{\cal E}_{\ell \ell}} - \sqrt{{\cal E}_{s s}} \right)^2.
\end{equation}

\vspace{0.1cm}

Eq.\,\eqref{eq:gammaStar} suggests that some fluids are a better choice than others for liquid-phase exfoliation because their surface energy is close to the surface energy of graphene. Indeed, it has been shown experimentally that exfoliation was the most efficient when performed with solvents such as N-methyl-pyrrolidone (NMP) and dimethylformamide (DMF) \cite{Hernandez2008, Coleman2013, Ravula2015}, whose surface energies are ${\cal E}_{\ell \ell} \sim 68$\,mJ/m$^2$. Therefore, Eq.\,\eqref{eq:gammaStar} suggests that the surface energy of graphene is ${\cal E}_{ss} \sim 68$\,mJ/m$^2$, a value that is reasonably close to that obtained with contact angle measurements \cite{Wang2009}. However, a broad range of values for the surface energy of graphene has been reported. For instance, a direct measurement of the surface energy using a surface force apparatus \cite{Babenko2017} gave ${\cal E}_{ss}=115 \pm 4$\,mJ/m$^2$. If we use this value in Eq.\,\eqref{eq:gammaStar}, we get $\dot \gamma_c=4 \cdot 10^6$\,s$^{-1}$ for micrometric particles in NMP fluid, which does not compare well with the experimental values of $\dot \gamma_c \approx 10^4$\,s$^{-1}$ \, \cite{Paton2014}. In addition, some solvent with surface energy ${\cal E}_{\ell \ell} \sim 68$\,mJ/m$^2$ are known to be a poor choice for graphene exfoliation \cite{Coleman2013}. Therefore, the high efficiency of NMP and DMF to exfoliate graphite nanoparticles remains a mystery, suggesting that the accuracy of Eq.\,\eqref{eq:gammaStar} has to be reconsidered.

\vspace{0.1cm}

It has been proposed that the Hansen solubility parameters, that accounts for dispersive, polar, and hydrogen-bonding components of the cohesive energy density of a material, is a much better indicator of the quality of a solvent for the exfoliation of graphene \cite{Hernandez2010, Coleman2013}. However, the Hansen solubility parameter also leads to contradictory results, as it suggests that ideal fluids for graphene dispersion are fluids with non-zero value of polar and hydrogen-bonding parameters, even though graphene is nonpolar \cite{Hernandez2010}.

\vspace{0.1cm}

In addition to experiments, molecular dynamics simulations have been used to evaluate the respective exfoliation efficiency of different fluids. Most authors have measured the potential of mean force (PMF) associated with the peeling of a layer, or the detachment of parallel rigid layers \cite{An2010, Shih2010, Sresht2015, Bordes2018a}. When performed in a liquid, such measurement gives precious information on the thermodynamic stability of dispersed graphene and have shown that NMP should have excellent performance for graphene exfoliation, in agreement with experimental data \cite{Bordes2018a}. However, PMF measurements are static and do not account for the dynamic effects associated with the exfoliation process. 

\vspace{0.1cm}

In this context, we perform out-of-equilibrium molecular dynamics (MD) simulations of the exfoliation of graphite platelets by different shearing fluids, starting with NMP and water. We record the critical shear rate $\dot \gamma_c$ above which exfoliation occurs, and compare our results with Eq.\,\eqref{eq:gammaStar}. Our results emphasis that Eq.\,\eqref{eq:gammaStar} is limited in its predictive capability and we, therefore, propose an alternative to Eq.\,\eqref{eq:gammaStar} that accounts for, among other effects, hydrodynamic slip. Slip reduces the hydrodynamic stress in the direction parallel to the surface and, therefore, significantly affects the tangential hydrodynamic force applied by the shearing liquid on the particle. 

\section{Result}

\noindent We perform MD simulations of a freely suspended graphite particle in a shear flow using LAMMPS \cite{LAMMPS}. The initial configuration consists of a stack of $N$ graphene layers immersed in a liquid, with $N$ between 2 and 6. Rigid walls are used to enclose the fluid in the $\vec{\mathrm e}_y$ direction (Fig.\,\ref{fig:FIGURE2}). Periodic boundary conditions are used in the three orthogonal directions. The effective thickness of the platelet in the $\vec{\mathrm e}_y$ direction is $H$, its length in the $\vec{\mathrm e}_x$ direction is $L$, and the span-wise dimension of the computational domain in the $\vec{\mathrm e}_z$ direction is $w$. The simulation box is equal to $3 \times L$ in the $\vec{\mathrm e}_x$ direction, $2$\,nm in the $\vec{\mathrm e}_y$ direction and the distance between the rigid walls is $H_w \approx 14$\,nm. Based on a preliminary convergence study, $H_w$ and the dimensions of the computational box were chosen large enough to avoid finite-size effects. We use the Adaptive Intermolecular Reactive Empirical Bond Order (AIREBO) force field for graphene \cite{Stuart2000}. The fluid consists of a number $N_f$ of water molecules or N-Methyl-2-pyrrolidone (NMP) molecules. We use the TIP4P/2005 model for water \cite{TIP4P} and the all-atom Gromos force field for NMP \cite{Schmid2011}. Carbon-fluid interaction parameters are calculated using the Lorentz-Berthelot mixing rule. The initial molecular structure of NMP is extracted from the automated topology builder \cite{Malde2011}. A shear flow of strength $\dot \gamma$ is produced by the relative translation of the two parallel walls in the $\vec{\mathrm e}_x$ direction, with respective velocity $U/2$ and $-U/2$. The two walls also impose atmospheric pressure on the fluid. Fluid molecules are maintained at a constant temperature $T=300$\,K using a Nos\'e-Hoover temperature thermostat \cite{Nose1984, Hoover1985} applied only to the degrees of freedom in the directions normal to the flow, $\vec{\mathrm e}_y$ and $\vec{\mathrm e}_z$.

\vspace{0.1cm}

During the initial stage of the simulation, the walls are allowed to move in the $\vec{\mathrm e}_y$ direction to impose a constant pressure of 1\,atm on the fluid, and the graphene layers are maintained immobile. After $50$\,ps, the graphene particle is allowed to freely translate and rotate using constant NVE integration, and the velocities of the walls in the $\vec{\mathrm e}_x$ direction are set equal to $U/2$ and $-U/2$ respectively, with $U=H_w \dot \gamma$. Each simulation is performed for a duration of $1$\,ns in addition to the $50$\,ps of the initial stage. Simulations are performed at fixed shear rate $\dot \gamma$, for a given number of layers $N$ and length $L$ of each layer. 

\vspace{0.1cm}

\begin{figure}
\includegraphics[width=\linewidth]{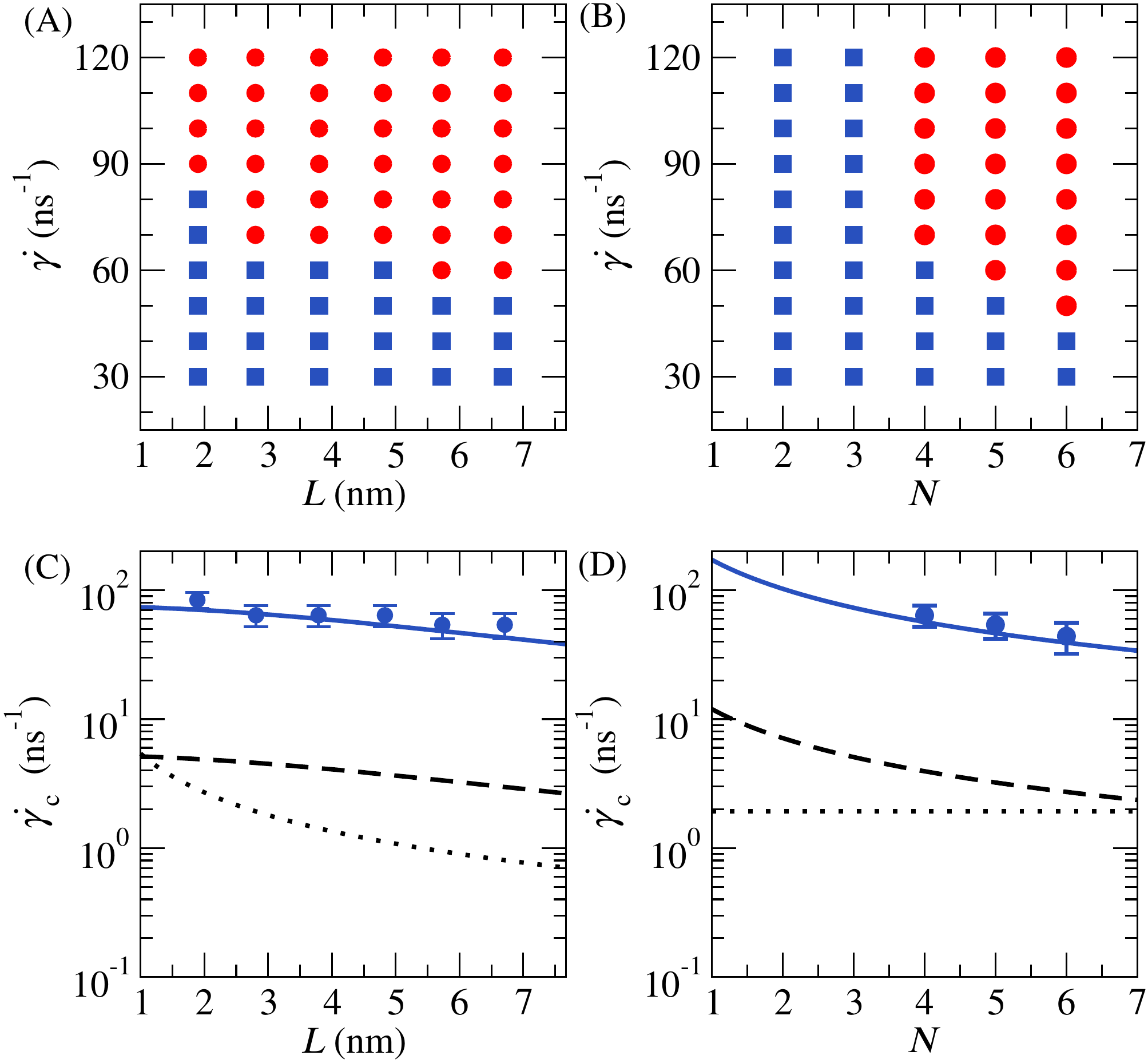}
\caption{\textbf{Exfoliation of graphite in water}
\textbf{(A)} Shear rate $\dot \gamma$ as a function of the nanoplatelet length $L$ for an initial number of layers $N = 4$. The red disks correspond to simulations for which exfoliation was observed, and blue squares to simulations for which exfoliation was not observed. 
\textbf{(B)} Shear rate $\dot \gamma$ as a function of $N$ for $L = 2.8$\,nm.
\textbf{(C)} Critical shear rate $\dot \gamma_c$ above which exfoliation occurs as a function of $L$ as extracted from MD simulation (symbols). Dotted black line is Eq.\,\eqref{eq:gammaStar}, dashed black line is Eq.\,\eqref{eq:gammaStar2}, and full blue line is Eq.\,\eqref{eq:gammaStar3}, see text for details.
\textbf{(D)} Critical shear rate $\dot \gamma_c$ as a function of the initial number of layers.
}
\label{fig:FIGURE3}
\end{figure}

The state of the graphite particle is controlled during the simulation, and two recurring situations are identified; (i) sliding of the layers does not occur (blue squares in Fig.\,\ref{fig:FIGURE3}\,A,\,B), or (ii) the platelet is exfoliated into a variable number of fewer-layer platelets (red disks in Fig.\,\ref{fig:FIGURE3}\,A,\,B). We associate the transition between the unaltered (blue) phase and the exfoliated (red) phase with a critical shear rate $\dot \gamma_c$; $\dot \gamma_c$ decreases with the nanoplatelet length, as well as with the initial number of nanoplatelet $N$ (Fig.\,\ref{fig:FIGURE3}\,C,\,D). In the case of water fluid, for an initial number of layer $N \le 3$, no exfoliation is observed, even for $\dot \gamma$ as high as 120\,ns$^{-1}$.

\vspace{0.1cm}

\begin{figure}
\includegraphics[width=\linewidth]{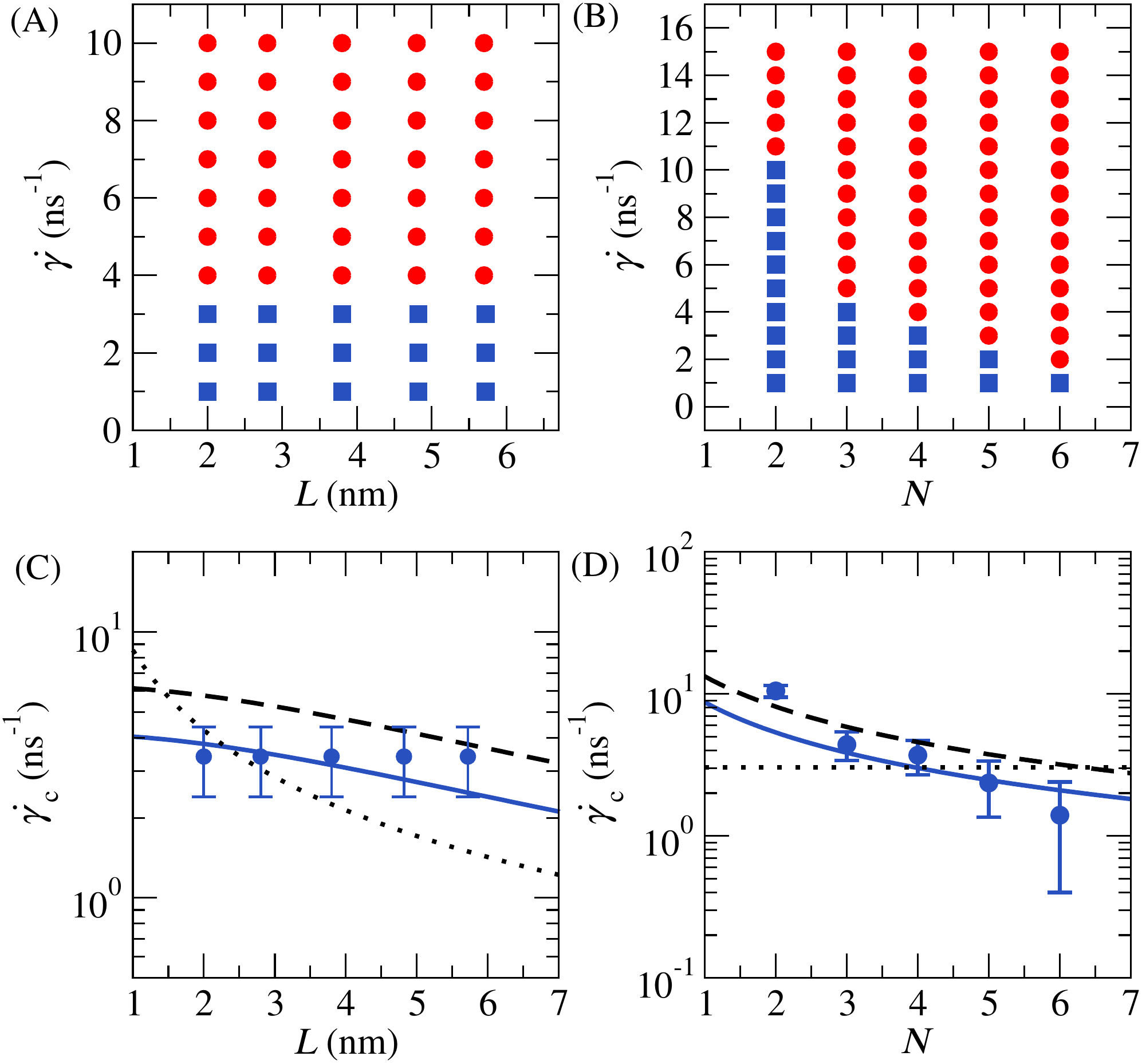}
\caption{\textbf{Exfoliation of graphite in NMP}
\textbf{(A)} Shear rate $\dot \gamma$ as a function of the nanoplatelet length $L$ for an initial number of platelet $N = 4$. The red disks correspond to simulations for which exfoliation was observed, and blue squares simulations for which exfoliation was not observed. 
\textbf{(B)} Shear rate $\dot \gamma$ as a function of $N$ and fixed $L = 2.8$\,nm.
\textbf{(C)} Critical shear rate $\dot \gamma_c$ above which exfoliation occurs as a function of $L$ as extracted from MD simulation (symbols). Dotted black line is Eq.\,\eqref{eq:gammaStar}, dashed black line is Eq.\,\eqref{eq:gammaStar2}, and full blue line is Eq.\,\eqref{eq:gammaStar3}, see text for details.
\textbf{(D)} Critical shear rate $\dot \gamma_c$ as a function of the initial number of layers.
}
\label{fig:FIGURE4}
\end{figure}

Similar simulations are performed using NMP (Fig.\,\ref{fig:FIGURE4}\,A,\,B). For NMP and a given number of layer $N$, the critical shear rate $\dot \gamma_c$ above which exfoliation is observed is typically one order of magnitude lower than in water (Fig.\,\ref{fig:FIGURE4}\,C,\,D), a difference that cannot be explained by the difference in viscosity of the two fluids ($\eta = 0.855$\,mPa\,s for TIP4P/2005 water at $300$\,K \cite{Gonzalez2010}, and $\eta = 1.6$\,mPa\,s for NMP \cite{Henni2004}). Unlike for water, in NMP fluid exfoliation is observed for any value of $N$ and $L$. 

\vspace{0.1cm}

The critical shear rate $\dot \gamma_c$ obtained using MD can be compared with the prediction of Eq.\,\eqref{eq:gammaStar}. To do so, both solid-solid ${\cal E}_{s s}$ and liquid-liquid ${\cal E}_{\ell \ell}$ surface energies are needed. The surface energy of graphene corresponds to half the work required to separate two initially bounded layers \cite{Henry2005}. We find ${\cal E}_{s s} = 147$\,mJ/m$^2$ for the AIREBO force field at zero temperature (Supporting Information). The liquid-liquid surface energy follows from the surface tension $\gamma$ as ${\cal E}_{\ell \ell} = \gamma + T S$, where $S$ is the entropy \cite{Ferguson1941}. Using the universal value for the entropy $S \sim 0.1$\,mJ m$^{-2}$K$^{-1}$ \, \cite{Paton2014}, and using literature values for the surface tension of water and NMP, one gets ${\cal E}_{\ell \ell} = 99.5$\,mJ/m$^2$ for water, and ${\cal E}_{\ell \ell} = 71$\,mJ/m$^2$ for NMP at $T=300$\,K \cite{Lopez2013, Alejandre2010}. Results show that Eq.\,\eqref{eq:gammaStar} fails to predict $\dot \gamma_c$, particularly in the case of water (dotted lines in Fig.\,\ref{fig:FIGURE3}\,C,\,D and Fig.\,\ref{fig:FIGURE4}\,C,\,D). In addition, Eq.\,\eqref{eq:gammaStar} predicts a functional form $\dot \gamma_c \propto L^{-1}$ that is in disagreement with the MD results, and fails to capture the variation of $\dot \gamma_c$ with the initial number of layers $N$. 

\section{Model for the exfoliation of nanoplatelet}

\noindent To improve the accuracy of Eq.\,\eqref{eq:gammaStar}, one first needs to improve the expression for the work of the hydrodynamic force, Eq.\,\eqref{eq:wf}. For nanomaterials with a smooth surface such as graphene and for most solvent, the classical no-slip boundary condition is often inaccurate and should be replaced by a partial slip boundary condition \cite{Kannam2013}. Hydrodynamic slip at the solid-liquid interface can be characterised by a Navier slip length $\lambda$, which is the distance within the solid at which the relative solid-fluid velocity extrapolates to zero \cite{Lauga2007, Bocquet2007}. In order to quantify the effect of slip on the hydrodynamic force, we consider the traction vector  $\boldsymbol{f} = \boldsymbol{\sigma} \cdot \boldsymbol{n}$, where $\boldsymbol{\sigma}$ is the fluid stress tensor and $\boldsymbol{n}$ is the normal to the surface. The traction can be calculated exactly by solving a boundary integral equation \cite{pozrikidis1992boundary}. For a thin particle aligned in the direction of the undisturbed shear flow  (at high shear rates an elongated particle spends most of the time aligned in the flow direction \cite{Singh2014,Kamal2019}), the traction can be estimated analytically by expanding the boundary integral equation to leading order in $H/L \ll 1$   \cite{Singh2014,Kamal2019}. Accounting for a Navier slip boundary condition \cite{pozrikidis1992boundary}, this asymptotic analysis yields the following leading-order expression for the hydrodynamic tangential traction \cite{Kamal2019}, valid far from the edges:
\begin{equation}
f_x \equiv \boldsymbol{f} \cdot \vec{e}_x=\dfrac{\dot \gamma \eta}{1+8 \lambda / (\pi L)}.
\end{equation}
Because $f_x$ is uniform, the leading-order contribution to $F_\text{hyd}$ from the flat surface of the graphene particle is
\begin{equation}
\label{eq:Fs0}
\int_{A_\text{t}} f_x \, \mathrm d S \approx \dfrac{\dot \gamma \eta w L}{1 + 8 \lambda/(\pi L)},
\end{equation}
where $|A_\text{t}| \approx w L$. Since the slip length for graphene in water is typically $\lambda \approx 10$\,nm \cite{Maali2008, Ortiz-Young2013, Tocci2014}, slip reduces the hydrodynamic force applied by the fluid on the platelet by a factor $1 + 8 \lambda / (\pi L) \approx 7$, assuming a length $L=4$\,nm.


\vspace{0.1cm}

\begin{figure}
\begin{center}
\includegraphics[width=\linewidth]{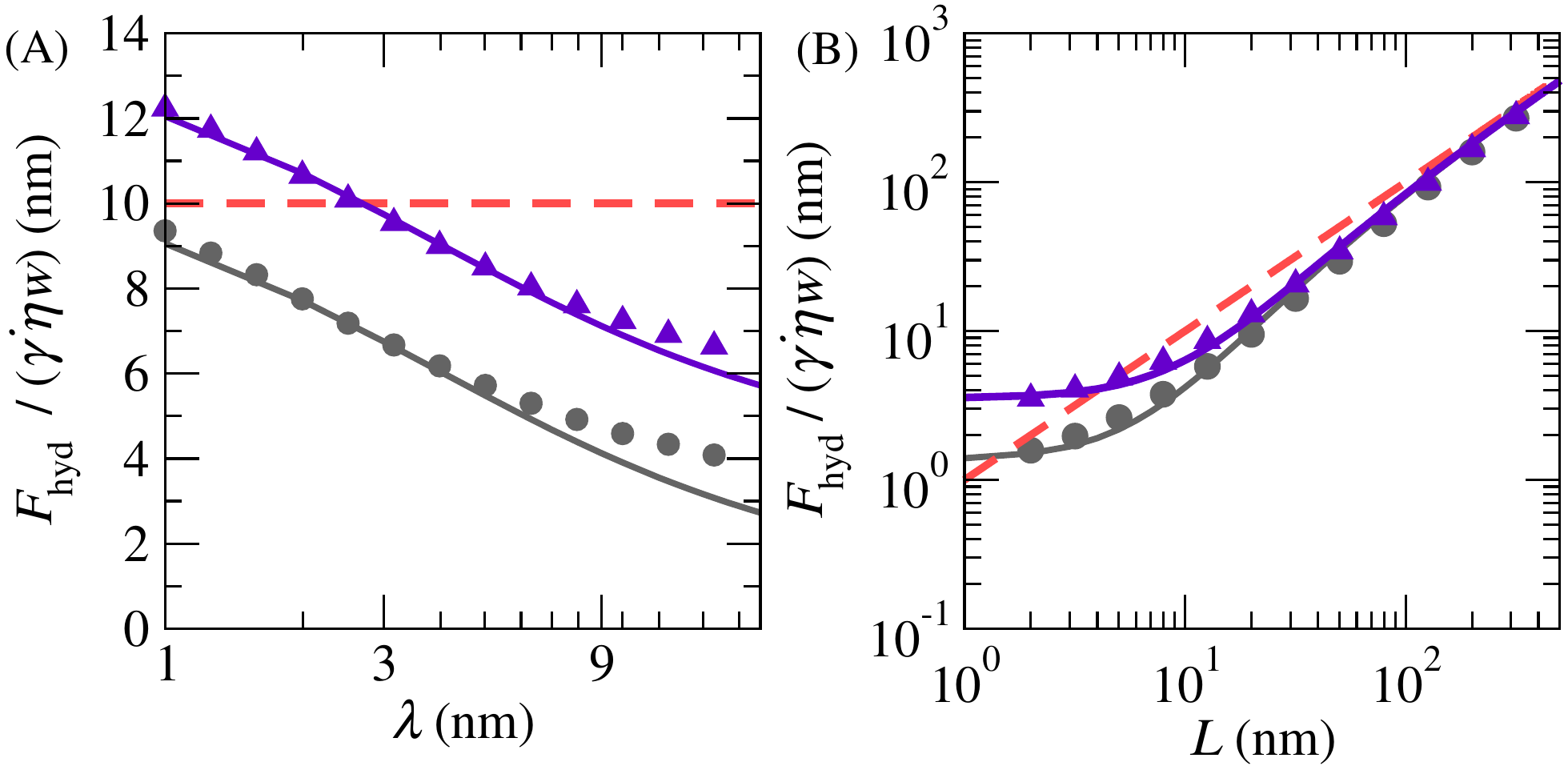}
\end{center}
\caption{
\textbf{(A)} Continuum calculations of the lateral hydrodynamic force applied by the liquid on the top platelet with length $L=10$\,nm, $N=2$ (grey disks), $N=8$ (purple triangles), as a function of the slip length $\lambda$. Calculations are made using the boundary integral method (BIM, Supporting Information). Full lines are Eq.\,\eqref{eq:Fs} with $c \approx 1.5$, and the red dashed line is the no-slip, no edge effect approximation, $F_\text{hyd} / (\dot \gamma \eta w) = L$.
\textbf{(B)} BIM calculations for a platelet with $\lambda=10$\,nm, $N=2$ (grey disks), $N=8$ (purple triangles), as a function of $L$. Full lines are Eq.\,\eqref{eq:Fs} with respectively $c=1.8$ ($N=2$) and $c=1.2$ ($N=8$), and the red dashed line is the no-slip, no edge effect approximation.}
\label{fig:FIGURE5}
\end{figure}

In addition to the force due to shearing of the flat surfaces, an additional hydrodynamic contribution is due to the force on the edges of the platelet \cite{Singh2014a}. For a nanometric platelet, these edge effects can even be dominant, in particular for a platelet with a large slip length \cite{Kamal2019}. 
Because  stresses in Stokes flow scale proportionally to $\eta \dot \gamma$, and the edge hydrodynamics is controlled by the thickness $H$, the edge force is expected to scale as $F_\text{hyd}^\text{e} \sim \dot \gamma \eta w  H$. Using the fact that $H \simeq N d$, where $d\approx 3.4$\,\AA\, is the inter-layer distance, we can write
\begin{equation}
\label{eq:Fse}
F_\text{hyd}^\text{e} \approx \dot \gamma \eta w c N d.
\end{equation}
Our simulation data from both the boundary integral method and MD simulations indeed supports the scaling of Eq.\,$\eqref{eq:Fse}$, suggesting $c \simeq 1.5$ (with some dispersion; actual values range between $1$ and $2$, suggesting a weak dependence on $N$ and  $\lambda$, see Fig.\,\ref{fig:FIGURE5}\,A,\,B, Fig.\,S1 Supporting Information). Including the edge force, the total hydrodynamic force driving inter-layer sliding  is 
\begin{equation}
\label{eq:Fs}
F_\text{hyd} \approx \dot \gamma \eta w \left( \dfrac{L}{1 + 8 \lambda/(\pi L)} + c N d \right).
\end{equation}
For $N=4$, $L=4$\,nm, and $\lambda=10$\,nm, one gets that the contribution from the edges (term containing $c N d$ in Eq.\,\eqref{eq:Fs}) is more than five times larger than the contribution from the flat surfaces (term containing $\lambda$ in Eq.\,\eqref{eq:Fs}). Not accounting for the corrections in Eq.\,\ref{eq:Fs} can lead to large errors, particularly for $L<20$\,nm (Fig.\,\ref{fig:FIGURE5}\,B).

\vspace{0.1cm}

Now inserting Eq.\,\eqref{eq:Fs} into the expression for the work (Eq.\,\eqref{eq:wf}) and balancing Eq.\,\eqref{eq:DE} and Eq.\,\eqref{eq:wf}, one gets a critical shear rate
\begin{equation}
\label{eq:gammaStar2}
\dot \gamma_c \approx \dfrac{1}{\eta} \dfrac{\left( \sqrt{{\cal E}_{\ell \ell}} - \sqrt{{\cal E}_{s s}} \right)^2}
{L / (1+8 \lambda / (\pi L)) + c N d}.
\end{equation}
Unlike Eq.\,\eqref{eq:gammaStar}, Eq.\,\eqref{eq:gammaStar2} appears to have the same trend as the MD data, for changes in $L$ or $N$ (dashed lines in Fig.\,\ref{fig:FIGURE3} and Fig.\,\ref{fig:FIGURE4}). Here we used our independent measurements for the slip length, respectively $\lambda=20$\,nm for water and $\lambda=12$\,nm for NMP (Supporting Information). However, predictions from Eq.\,\eqref{eq:gammaStar2} are still in quantitative disagreement with the MD measurements, suggesting that the remaining problem is in the estimation of $\Delta E$. 

\vspace{0.1cm}

Eq.\,\eqref{eq:DE} has been obtained using the geometric mixing rule. However, this semi-empirical rule is not accurate in general for predicting solid-liquid surface energy, in particular for fluids with a polar contribution \cite{Shen2015}. To prove this, instead of using the mixing rule, we retain all three energy terms and evaluate $\Delta E$ from Eq.\,\eqref{eq:DEfull}, leading to 
\begin{equation}
\label{eq:gammaStar3}
\dot \gamma_c \approx \dfrac{1}{\eta} \dfrac{ {\cal E}_{\ell \ell} + {\cal E}_{s s} - 2 {\cal E}_{\ell s}}
{L / (1+8 \lambda / (\pi L)) + c N d}.
\end{equation}
The agreement between Eq.\,\eqref{eq:gammaStar3} and MD is excellent using $c \approx 1$, ${\cal E}_{\ell s} = 90$\,mJ/m$^2$ for water, and ${\cal E}_{\ell s} = 105$\,mJ/m$^2$ for NMP. These surface energy density values are very close to those we obtained by performing independent measurements of ${\cal E}_{\ell s}$ obtained by measuring the difference between longitudinal and transverse pressures near a fluid-solid or fluid-vapour interface \cite{Kirkwood1949,Dreher2018a, Dreher2019b}. We found ${\cal E}_{\ell s} = 93 \pm 4$\,mJ/m$^2$ for water-graphene, and ${\cal E}_{\ell s} = 107 \pm 5$\,mJ/m$^2$ for NMP-graphene (Supporting Information). 

\vspace{0.1cm}

\begin{figure}
\begin{center}
\includegraphics[width=\linewidth]{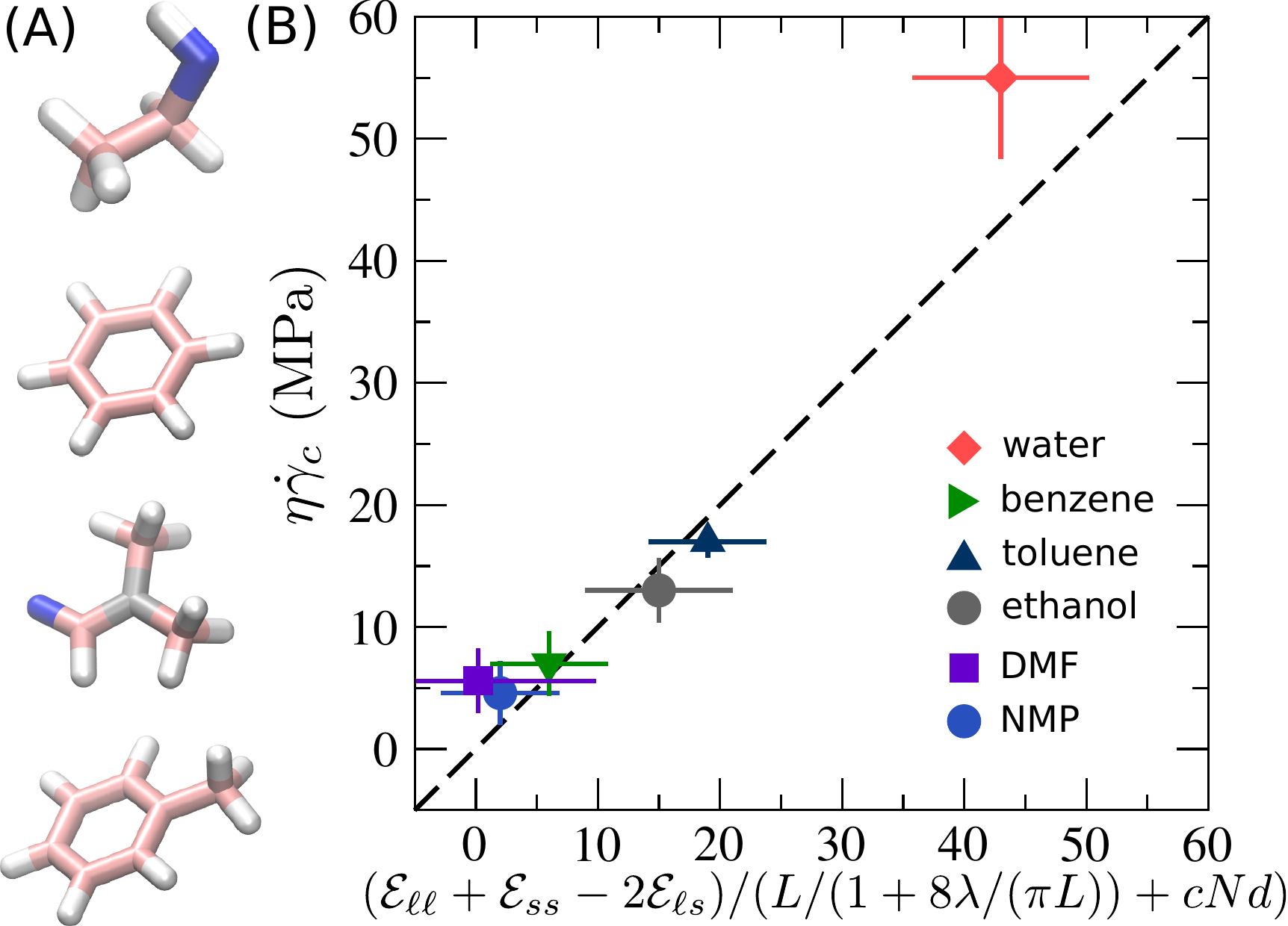}
\end{center}
\caption{\textbf{(A)} From top to bottom: ethanol, benzene, DMF, and toluene molecules. \textbf{(B)} Critical shear stress $\eta \dot \gamma_c$ as measured from MD as a function of the normalised energy difference $({\cal E}_{\ell \ell} + {\cal E}_{s s} - 2 {\cal E}_{\ell s}) / ( L / (1+8 \lambda / (\pi L)) + c N d)$ (expressed in mJ/m$^2$/nm) with $c \approx 1$, see the text for details. Dashed line is a $y=x$ guide for the eyes.}
\label{fig:FIGURE6}
\end{figure}

To test further the general applicability to different solvents of Eq.\,\eqref{eq:gammaStar3}, we performed simulations using four additional liquids: ethanol, benzene, DMF, and toluene (Fig.\,\ref{fig:FIGURE6}\,A). These solvents have been selected for their low viscosity ($ \le 1.1$\,mPa\,s), and because together with water and NMP, they offer a broad range of surface energy values (Table S2, Supporting Information). For a graphite particle of length $L=3.6$\,nm and initial layer number $N=4$, we have extracted the critical shear rate $\dot \gamma_c$ for each solvent. We report the critical shear stress values $\eta \dot \gamma_c$ for each fluid as a function of $({\cal E}_{\ell \ell} + {\cal E}_{s s} - 2 {\cal E}_{\ell s}) / (L / (1+8 \lambda / (\pi L)) + c N d)$, where the surface energy ${\cal E}_{\ell s}$ and slip length have been measured independently for each fluid. MD results for the seven different fluids show a good agreement with Eq.\,\eqref{eq:gammaStar3} (Fig.\,\ref{fig:FIGURE6}\,B).

\section{Discussion}

\noindent In this article, we used out of equilibrium MD to simulate the exfoliation of defect-free graphite nanoplatelet. We measured the critical shear rate $\dot \gamma_c$ above which exfoliation occurs using different solvents, with a particular focus on comparing NMP, typically considered an optimal solvent for the exfoliation of pure graphene, and water, typically considered not a good solvent. We compared the MD results with a simple theoretical model based on a balance between the work done by hydrodynamic forces and the change in interfacial energy associated with the separation of the layers. We find a good agreement between the model and MD provided that (i) the hydrodynamics force accounts for slip at the solid-fluid interface, (ii) the hydrodynamics force accounts for additional edge-related contributions, (iii) and that the full energy difference associated with the separation of the layers is accounted for.

\vspace{0.1cm}

Since the validity of Eq.\,\eqref{eq:gammaStar3} has been demonstrated by comparison with MD, we can use it to predict the critical shear rate $\dot \gamma_c$ for a platelet with more realistic dimensions, and compare the outcome with experimental data. Using microfluidisation, Karagiannidis et al.\cite{Karagiannidis2017} have reported the exfoliation of graphite in aqueous solution (sodium deoxycholate and deionized water) for shear rates above $\dot \gamma_c \sim 10^8$\,s$^{-1}$. Assuming $L=1$\,\si{\micro}m, the mean flake size reported by Karagiannidis et al., as well as $N=10$, and $\lambda = 10$\,nm, a typical experimental slip length value for graphene \cite{Maali2008, Ortiz-Young2013}, we have $\lambda/L \ll 1 $ and $c N d/L \ll 1$ (Fig\,\ref{fig:FIGURE5}\,B) such that 
\begin{equation}
\label{eq:gammaStar4}
\dot \gamma_c \approx \dfrac{1}{\eta L} ( {\cal E}_{\ell \ell} + {\cal E}_{s s} - 2 {\cal E}_{\ell s}).
\end{equation}
Using the MD's values for ${\cal E}_{s s}$ and ${\cal E}_{\ell s}$, together with the experimental values for $\eta$ and ${\cal E}_{\ell \ell}$ (table S2, Supporting Information), Eq.\,\eqref{eq:gammaStar4} gives $\dot \gamma_c = 6 \cdot 10^7$\,s$^{-1}$, which is close to the value of $\dot \gamma_c \sim 10^8$\,s$^{-1}$ reported by Karagiannidis et al. \cite{Karagiannidis2017}. 

\vspace{0.1cm}

Using a rotating mixer, Paton et al. have reported the exfoliation of graphite in NMP for shear rates above $\dot \gamma_c \sim 10^4$\,s$^{-1}$. In such experiments, $\lambda/L \ll 1$ and $c N d/L \ll 1$, so Eq.\,\eqref{eq:gammaStar4} can again be used to predict the critical shear rate. Using the MD's values for ${\cal E}_{s s}$ and ${\cal E}_{\ell s}$, together with the experimental values for $\eta$ and ${\cal E}_{\ell \ell}$, Eq.\,\eqref{eq:gammaStar4} predicts $\dot \gamma_c = 4 \cdot 10^6$\,s$^{-1}$, a value that is two orders of magnitude larger than the experimental value. 

\vspace{0.1cm}

There are many potential reasons for this discrepancy. One is the sensitivity of the model parameters. A sensitivity analysis of Eq.\,\ref{eq:gammaStar4} can be carried out by first writing
\begin{equation}
\dot \gamma_c^\text{o} \approx \dfrac{1}{\eta^\text{o} L} ( {\cal E}_{\ell \ell}^\text{o} + {\cal E}_{s s}^\text{o} - 2 {\cal E}_{\ell s}^\text{o}),
\end{equation}
where the superscript `o' refers to the \textit{observed} experimental parameters, and then comparing this expression with Eq.\,\eqref{eq:gammaStar4}, which contained \textit{estimated} parameters (from MD). Assuming that the only uncertainties are in the value of ${\cal E}_{\ell s}$ (i.e. ${\cal E}_{\ell \ell}^\text{o} = {\cal E}_{\ell \ell}$, ${\cal E}_{s s}^\text{o} = {\cal E}_{s s}$ and $\eta^\text{o}=\eta$), one can write the difference between the observed critical shear rate and the predicted one as
\begin{equation}
\dot \gamma_c - \dot \gamma_c^\text{o} \approx \dfrac{2}{\eta L} ({\cal E}_{\ell s}^\text{o} - {\cal E}_{\ell s}).
\end{equation}
Assuming a $1\%$ difference between ${\cal E}_{\ell s}$ and ${\cal E}_{\ell s}^\text{o}$, and since ${\cal E}_{\ell s}^\text{o}$ is typically of the order of $100$\,mJ/m$^2$, one gets $\dot \gamma_c - \dot \gamma_c^\text{o}$ of the order of $10^6$\,s$^{-1}$ for $L=1$\,\si{\micro}m and $\eta=1$\,mPa\,s. Since in the case of NMP, $\dot \gamma_c^\text{o} \approx 10^{4}$\,s$^{-1}$, an error of only one percent on ${\cal E}_{\ell s}$ leads to a difference by two orders of magnitude between the predicted and observed value of $\dot \gamma_c$.
This analysis demonstrates the challenge of drawing definite conclusions regarding the validity of the model by comparing it against experiments in which surface energy parameters are not measured independently.

\vspace{0.1cm}

A second explanation for the discrepancy is the possible importance in experiments of bending deformations. Bending deformations are relatively unimportant in our MD simulations because the nanosheets have small lengths and are therefore relatively rigid, but the same cannot be said for micro and nanosheets having $L$ in the micron range. For graphene multilayers where at least one of the layers deforms significantly by bending, the energy balance should include a bending energy term associated with the internal work of deformation of the solid, in addition to external work and adhesion energy terms \cite{Lingard1974}. Simple dimensional analysis suggests that the most general expression for the critical shear rate is \cite{botto2019towards}
\begin{equation}
\dot{\gamma}_c \approx \frac{1}{\eta L} \frac{{\cal E}_{\ell \ell} + {\cal E}_{s s} - 2 {\cal E}_{\ell s}}{g(\dot{\gamma} \eta L^3/B)}
\end{equation}
where $g$ is a non-dimensional function that accounts for the effect of flexibility on the force resisting exfoliation (e.g. accounting for stress concentration effects in peeling deformations), and $B$ is the bending rigidity of the deforming layer. For $\dot{\gamma} \eta L^3/B \ll 1$ (rigid sheets), $g$ is expected to tend to 1, recovering Eq.\,\eqref{eq:gammaStar4}. However, for $\dot{\gamma} \eta L^3/B \sim 1$ or larger, bending deformations become important, and a stronger dependence of $\dot{\gamma}$ on $L$ emerges \cite{botto2019towards, Salussolia2019}. The considerations made in this paper regarding the quantification of surface energies and hydrodynamic force contributions, however, remain valid and the comparison of Eq.\,\eqref{eq:gammaStar3} with MD is an important stepping stone towards accurate predictive models of exfoliation. 

\section*{Supporting Information}

\noindent I) Energy measurement at solid interfaces. II) Slip length measurements. III) Hydrodynamic force measurement. IV) Parameters for the 6 fluids. Figure S1. Table S2. 

\section*{Acknowledgements}

\noindent The authors thank the European Research Council (ERC) for funding towards the project {\scshape flexnanoflow} (n$^\circ\,715475$). This research utilised Queen Mary's Apocrita HPC facility, supported by QMUL Research-IT. 

\bibliography{Simon.bib}

\begin{thebibliography}{53}%
\makeatletter
\providecommand \@ifxundefined [1]{%
 \@ifx{#1\undefined}
}%
\providecommand \@ifnum [1]{%
 \ifnum #1\expandafter \@firstoftwo
 \else \expandafter \@secondoftwo
 \fi
}%
\providecommand \@ifx [1]{%
 \ifx #1\expandafter \@firstoftwo
 \else \expandafter \@secondoftwo
 \fi
}%
\providecommand \natexlab [1]{#1}%
\providecommand \enquote  [1]{``#1''}%
\providecommand \bibnamefont  [1]{#1}%
\providecommand \bibfnamefont [1]{#1}%
\providecommand \citenamefont [1]{#1}%
\providecommand \href@noop [0]{\@secondoftwo}%
\providecommand \href [0]{\begingroup \@sanitize@url \@href}%
\providecommand \@href[1]{\@@startlink{#1}\@@href}%
\providecommand \@@href[1]{\endgroup#1\@@endlink}%
\providecommand \@sanitize@url [0]{\catcode `\\12\catcode `\$12\catcode
  `\&12\catcode `\#12\catcode `\^12\catcode `\_12\catcode `\%12\relax}%
\providecommand \@@startlink[1]{}%
\providecommand \@@endlink[0]{}%
\providecommand \url  [0]{\begingroup\@sanitize@url \@url }%
\providecommand \@url [1]{\endgroup\@href {#1}{\urlprefix }}%
\providecommand \urlprefix  [0]{URL }%
\providecommand \Eprint [0]{\href }%
\providecommand \doibase [0]{http://dx.doi.org/}%
\providecommand \selectlanguage [0]{\@gobble}%
\providecommand \bibinfo  [0]{\@secondoftwo}%
\providecommand \bibfield  [0]{\@secondoftwo}%
\providecommand \translation [1]{[#1]}%
\providecommand \BibitemOpen [0]{}%
\providecommand \bibitemStop [0]{}%
\providecommand \bibitemNoStop [0]{.\EOS\space}%
\providecommand \EOS [0]{\spacefactor3000\relax}%
\providecommand \BibitemShut  [1]{\csname bibitem#1\endcsname}%
\let\auto@bib@innerbib\@empty
\bibitem [{\citenamefont {Mas-Ballest{\'{e}}}\ \emph
  {et~al.}(2011)\citenamefont {Mas-Ballest{\'{e}}}, \citenamefont
  {G{\'{o}}mez-Navarro}, \citenamefont {G{\'{o}}mez-Herrero},\ and\
  \citenamefont {Zamora}}]{Mas-Balleste2011}%
  \BibitemOpen
  \bibfield  {author} {\bibinfo {author} {\bibfnamefont {R.}~\bibnamefont
  {Mas-Ballest{\'{e}}}}, \bibinfo {author} {\bibfnamefont {C.}~\bibnamefont
  {G{\'{o}}mez-Navarro}}, \bibinfo {author} {\bibfnamefont {J.}~\bibnamefont
  {G{\'{o}}mez-Herrero}}, \ and\ \bibinfo {author} {\bibfnamefont
  {F.}~\bibnamefont {Zamora}},\ }\bibfield  {title} {\enquote {\bibinfo {title}
  {{2D materials: To graphene and beyond}},}\ }\href {\doibase
  10.1039/c0nr00323a} {\bibfield  {journal} {\bibinfo  {journal} {Nanoscale}\
  }\textbf {\bibinfo {volume} {3}},\ \bibinfo {pages} {20--30} (\bibinfo {year}
  {2011})}\BibitemShut {NoStop}%
\bibitem [{\citenamefont {Mounet}\ \emph {et~al.}(2018)\citenamefont {Mounet},
  \citenamefont {Gibertini}, \citenamefont {Schwaller}, \citenamefont {Campi},
  \citenamefont {Merkys}, \citenamefont {Marrazzo}, \citenamefont {Sohier},
  \citenamefont {Castelli}, \citenamefont {Cepellotti}, \citenamefont {Pizzi},\
  and\ \citenamefont {Marzari}}]{Mounet2018}%
  \BibitemOpen
  \bibfield  {author} {\bibinfo {author} {\bibfnamefont {N.}~\bibnamefont
  {Mounet}}, \bibinfo {author} {\bibfnamefont {M.}~\bibnamefont {Gibertini}},
  \bibinfo {author} {\bibfnamefont {P.}~\bibnamefont {Schwaller}}, \bibinfo
  {author} {\bibfnamefont {D.}~\bibnamefont {Campi}}, \bibinfo {author}
  {\bibfnamefont {A.}~\bibnamefont {Merkys}}, \bibinfo {author} {\bibfnamefont
  {A.}~\bibnamefont {Marrazzo}}, \bibinfo {author} {\bibfnamefont
  {T.}~\bibnamefont {Sohier}}, \bibinfo {author} {\bibfnamefont {I.~E.}\
  \bibnamefont {Castelli}}, \bibinfo {author} {\bibfnamefont {A.}~\bibnamefont
  {Cepellotti}}, \bibinfo {author} {\bibfnamefont {G.}~\bibnamefont {Pizzi}}, \
  and\ \bibinfo {author} {\bibfnamefont {N.}~\bibnamefont {Marzari}},\
  }\bibfield  {title} {\enquote {\bibinfo {title} {{Two-dimensional materials
  from high-throughput computational exfoliation of experimentally known
  compounds}},}\ }\href {\doibase 10.1038/s41565-017-0035-5} {\bibfield
  {journal} {\bibinfo  {journal} {Nature Nanotechnology}\ }\textbf {\bibinfo
  {volume} {13}},\ \bibinfo {pages} {246--252} (\bibinfo {year}
  {2018})}\BibitemShut {NoStop}%
\bibitem [{\citenamefont {Butler}\ \emph {et~al.}(2013)\citenamefont {Butler},
  \citenamefont {Hollen}, \citenamefont {Cao}, \citenamefont {Cui},
  \citenamefont {Gupta}, \citenamefont {Guti{\'{e}}rrez}, \citenamefont
  {Heinz}, \citenamefont {Hong}, \citenamefont {Huang}, \citenamefont {Ismach},
  \citenamefont {Johnston-Halperin}, \citenamefont {Kuno}, \citenamefont
  {Plashnitsa}, \citenamefont {Robinson}, \citenamefont {Ruoff}, \citenamefont
  {Salahuddin}, \citenamefont {Shan}, \citenamefont {Shi}, \citenamefont
  {Spencer}, \citenamefont {Terrones}, \citenamefont {Windl},\ and\
  \citenamefont {Goldberger}}]{Butler2013}%
  \BibitemOpen
  \bibfield  {author} {\bibinfo {author} {\bibfnamefont {S.~Z.}\ \bibnamefont
  {Butler}}, \bibinfo {author} {\bibfnamefont {S.~M.}\ \bibnamefont {Hollen}},
  \bibinfo {author} {\bibfnamefont {L.}~\bibnamefont {Cao}}, \bibinfo {author}
  {\bibfnamefont {Y.}~\bibnamefont {Cui}}, \bibinfo {author} {\bibfnamefont
  {J.~A.}\ \bibnamefont {Gupta}}, \bibinfo {author} {\bibfnamefont {H.~R.}\
  \bibnamefont {Guti{\'{e}}rrez}}, \bibinfo {author} {\bibfnamefont {T.~F.}\
  \bibnamefont {Heinz}}, \bibinfo {author} {\bibfnamefont {S.~S.}\ \bibnamefont
  {Hong}}, \bibinfo {author} {\bibfnamefont {J.}~\bibnamefont {Huang}},
  \bibinfo {author} {\bibfnamefont {A.~F.}\ \bibnamefont {Ismach}}, \bibinfo
  {author} {\bibfnamefont {E.}~\bibnamefont {Johnston-Halperin}}, \bibinfo
  {author} {\bibfnamefont {M.}~\bibnamefont {Kuno}}, \bibinfo {author}
  {\bibfnamefont {V.~V.}\ \bibnamefont {Plashnitsa}}, \bibinfo {author}
  {\bibfnamefont {R.~D.}\ \bibnamefont {Robinson}}, \bibinfo {author}
  {\bibfnamefont {R.~S.}\ \bibnamefont {Ruoff}}, \bibinfo {author}
  {\bibfnamefont {S.}~\bibnamefont {Salahuddin}}, \bibinfo {author}
  {\bibfnamefont {J.}~\bibnamefont {Shan}}, \bibinfo {author} {\bibfnamefont
  {L.}~\bibnamefont {Shi}}, \bibinfo {author} {\bibfnamefont {M.~G.}\
  \bibnamefont {Spencer}}, \bibinfo {author} {\bibfnamefont {M.}~\bibnamefont
  {Terrones}}, \bibinfo {author} {\bibfnamefont {W.}~\bibnamefont {Windl}}, \
  and\ \bibinfo {author} {\bibfnamefont {J.~E.}\ \bibnamefont {Goldberger}},\
  }\bibfield  {title} {\enquote {\bibinfo {title} {{Progress, challenges, and
  opportunities in two-dimensional materials beyond graphene}},}\ }\href
  {\doibase 10.1021/nn400280c} {\bibfield  {journal} {\bibinfo  {journal} {ACS
  Nano}\ }\textbf {\bibinfo {volume} {7}},\ \bibinfo {pages} {2898--2926}
  (\bibinfo {year} {2013})}\BibitemShut {NoStop}%
\bibitem [{\citenamefont {Geim}(2010)}]{Geim2010}%
  \BibitemOpen
  \bibfield  {author} {\bibinfo {author} {\bibfnamefont {A.~K.}\ \bibnamefont
  {Geim}},\ }\bibfield  {title} {\enquote {\bibinfo {title} {{Graphene : Status
  and Prospects}},}\ }\href {\doibase 10.1126/science.1158877} {\bibfield
  {journal} {\bibinfo  {journal} {Science}\ }\textbf {\bibinfo {volume}
  {1530}},\ \bibinfo {pages} {1530--1535} (\bibinfo {year} {2010})}\BibitemShut
  {NoStop}%
\bibitem [{\citenamefont {Avouris}\ and\ \citenamefont
  {Xia}(2012)}]{Avouris2012}%
  \BibitemOpen
  \bibfield  {author} {\bibinfo {author} {\bibfnamefont {P.}~\bibnamefont
  {Avouris}}\ and\ \bibinfo {author} {\bibfnamefont {F.}~\bibnamefont {Xia}},\
  }\bibfield  {title} {\enquote {\bibinfo {title} {{Graphene applications in
  electronics and photonics}},}\ }\href {\doibase 10.1557/mrs.2012.206}
  {\bibfield  {journal} {\bibinfo  {journal} {Graphene Fundamentals and
  Functionalities}\ }\textbf {\bibinfo {volume} {37}},\ \bibinfo {pages}
  {1225--1234} (\bibinfo {year} {2012})}\BibitemShut {NoStop}%
\bibitem [{\citenamefont {Brownson}, \citenamefont {Kampouris},\ and\
  \citenamefont {Banks}(2011)}]{Brownson2011}%
  \BibitemOpen
  \bibfield  {author} {\bibinfo {author} {\bibfnamefont {D.~A.~C.}\
  \bibnamefont {Brownson}}, \bibinfo {author} {\bibfnamefont {D.~K.}\
  \bibnamefont {Kampouris}}, \ and\ \bibinfo {author} {\bibfnamefont {C.~E.}\
  \bibnamefont {Banks}},\ }\bibfield  {title} {\enquote {\bibinfo {title} {{An
  overview of graphene in energy production and storage applications}},}\
  }\href {\doibase 10.1016/j.jpowsour.2011.02.022} {\bibfield  {journal}
  {\bibinfo  {journal} {Journal of Power Sources}\ }\textbf {\bibinfo {volume}
  {196}},\ \bibinfo {pages} {4873--4885} (\bibinfo {year} {2011})}\BibitemShut
  {NoStop}%
\bibitem [{\citenamefont {Chung}\ \emph {et~al.}(2013)\citenamefont {Chung},
  \citenamefont {Kim}, \citenamefont {Shin}, \citenamefont {Ryoo},
  \citenamefont {Hong},\ and\ \citenamefont {Min}}]{Chung2013}%
  \BibitemOpen
  \bibfield  {author} {\bibinfo {author} {\bibfnamefont {C.}~\bibnamefont
  {Chung}}, \bibinfo {author} {\bibfnamefont {Y.-k.}\ \bibnamefont {Kim}},
  \bibinfo {author} {\bibfnamefont {D.}~\bibnamefont {Shin}}, \bibinfo {author}
  {\bibfnamefont {S.-r.}\ \bibnamefont {Ryoo}}, \bibinfo {author}
  {\bibfnamefont {B.~H. E.~E.}\ \bibnamefont {Hong}}, \ and\ \bibinfo {author}
  {\bibfnamefont {D.-h.}\ \bibnamefont {Min}},\ }\bibfield  {title} {\enquote
  {\bibinfo {title} {{Biomedical Applications of Graphene and Graphene
  Oxide}},}\ }\href@noop {} {\bibfield  {journal} {\bibinfo  {journal}
  {Accounts of chemical research}\ }\textbf {\bibinfo {volume} {46}},\ \bibinfo
  {pages} {2211--2224} (\bibinfo {year} {2013})}\BibitemShut {NoStop}%
\bibitem [{\citenamefont {Yi}\ and\ \citenamefont {Shen}(2015)}]{Yi2015}%
  \BibitemOpen
  \bibfield  {author} {\bibinfo {author} {\bibfnamefont {M.}~\bibnamefont
  {Yi}}\ and\ \bibinfo {author} {\bibfnamefont {Z.}~\bibnamefont {Shen}},\
  }\bibfield  {title} {\enquote {\bibinfo {title} {{A review on mechanical
  exfoliation for scalable production of graphene}},}\ }\href {\doibase
  10.1039/x0xx00000x} {\bibfield  {journal} {\bibinfo  {journal} {Journal of
  Materials Chemistry A}\ }\textbf {\bibinfo {volume} {3}} (\bibinfo {year}
  {2015}),\ 10.1039/x0xx00000x}\BibitemShut {NoStop}%
\bibitem [{\citenamefont {Hernandez}\ \emph {et~al.}(2008)\citenamefont
  {Hernandez}, \citenamefont {Nicolosi}, \citenamefont {Lotya}, \citenamefont
  {Blighe}, \citenamefont {Sun}, \citenamefont {De}, \citenamefont {Mcgovern},
  \citenamefont {Holland}, \citenamefont {Byrne}, \citenamefont {Ko},
  \citenamefont {Boland}, \citenamefont {Niraj}, \citenamefont {Duesberg},
  \citenamefont {Krishnamurthy}, \citenamefont {Goodhue}, \citenamefont
  {Hutchison}, \citenamefont {Scardaci}, \citenamefont {Ferrari},\ and\
  \citenamefont {Coleman}}]{Hernandez2008}%
  \BibitemOpen
  \bibfield  {author} {\bibinfo {author} {\bibfnamefont {Y.}~\bibnamefont
  {Hernandez}}, \bibinfo {author} {\bibfnamefont {V.}~\bibnamefont {Nicolosi}},
  \bibinfo {author} {\bibfnamefont {M.}~\bibnamefont {Lotya}}, \bibinfo
  {author} {\bibfnamefont {F.~M.}\ \bibnamefont {Blighe}}, \bibinfo {author}
  {\bibfnamefont {Z.}~\bibnamefont {Sun}}, \bibinfo {author} {\bibfnamefont
  {S.}~\bibnamefont {De}}, \bibinfo {author} {\bibfnamefont {I.~T.}\
  \bibnamefont {Mcgovern}}, \bibinfo {author} {\bibfnamefont {B.}~\bibnamefont
  {Holland}}, \bibinfo {author} {\bibfnamefont {M.}~\bibnamefont {Byrne}},
  \bibinfo {author} {\bibfnamefont {Y.~K. G. U.~N.}\ \bibnamefont {Ko}},
  \bibinfo {author} {\bibfnamefont {J.~J.}\ \bibnamefont {Boland}}, \bibinfo
  {author} {\bibfnamefont {P.}~\bibnamefont {Niraj}}, \bibinfo {author}
  {\bibfnamefont {G.}~\bibnamefont {Duesberg}}, \bibinfo {author}
  {\bibfnamefont {S.}~\bibnamefont {Krishnamurthy}}, \bibinfo {author}
  {\bibfnamefont {R.}~\bibnamefont {Goodhue}}, \bibinfo {author} {\bibfnamefont
  {J.}~\bibnamefont {Hutchison}}, \bibinfo {author} {\bibfnamefont
  {V.}~\bibnamefont {Scardaci}}, \bibinfo {author} {\bibfnamefont {A.~C.}\
  \bibnamefont {Ferrari}}, \ and\ \bibinfo {author} {\bibfnamefont {J.~N.}\
  \bibnamefont {Coleman}},\ }\bibfield  {title} {\enquote {\bibinfo {title}
  {{High-yield production of graphene by liquid-phase exfoliation of
  graphite}},}\ }\href {\doibase 10.1038/nnano.2008.215} {\bibfield  {journal}
  {\bibinfo  {journal} {Nature nanotechnology}\ }\textbf {\bibinfo {volume}
  {3}},\ \bibinfo {pages} {563--568} (\bibinfo {year} {2008})}\BibitemShut
  {NoStop}%
\bibitem [{\citenamefont {Coleman}(2013)}]{Coleman2013}%
  \BibitemOpen
  \bibfield  {author} {\bibinfo {author} {\bibfnamefont {J.~N.}\ \bibnamefont
  {Coleman}},\ }\bibfield  {title} {\enquote {\bibinfo {title} {{Liquid
  exfoliation of defect-free graphene}},}\ }\href {\doibase 10.1021/ar300009f}
  {\bibfield  {journal} {\bibinfo  {journal} {Accounts of Chemical Research}\
  }\textbf {\bibinfo {volume} {46}},\ \bibinfo {pages} {14--22} (\bibinfo
  {year} {2013})}\BibitemShut {NoStop}%
\bibitem [{\citenamefont {Yi}\ and\ \citenamefont {Shen}(2014)}]{Yi2014}%
  \BibitemOpen
  \bibfield  {author} {\bibinfo {author} {\bibfnamefont {M.}~\bibnamefont
  {Yi}}\ and\ \bibinfo {author} {\bibfnamefont {Z.}~\bibnamefont {Shen}},\
  }\bibfield  {title} {\enquote {\bibinfo {title} {{Kitchen blender for
  producing high-quality few-layer graphene}},}\ }\href {\doibase
  10.1016/j.carbon.2014.07.035} {\bibfield  {journal} {\bibinfo  {journal}
  {Carbon}\ }\textbf {\bibinfo {volume} {78}},\ \bibinfo {pages} {622--626}
  (\bibinfo {year} {2014})}\BibitemShut {NoStop}%
\bibitem [{\citenamefont {Chen}, \citenamefont {Dobson},\ and\ \citenamefont
  {Raston}(2012)}]{Chen2012}%
  \BibitemOpen
  \bibfield  {author} {\bibinfo {author} {\bibfnamefont {X.}~\bibnamefont
  {Chen}}, \bibinfo {author} {\bibfnamefont {J.~F.}\ \bibnamefont {Dobson}}, \
  and\ \bibinfo {author} {\bibfnamefont {C.~L.}\ \bibnamefont {Raston}},\
  }\bibfield  {title} {\enquote {\bibinfo {title} {{Vortex fluidic exfoliation
  of graphite and boron nitride}},}\ }\href {\doibase 10.1039/c2cc17611d}
  {\bibfield  {journal} {\bibinfo  {journal} {Chemical Communications}\
  }\textbf {\bibinfo {volume} {48}},\ \bibinfo {pages} {3703--3705} (\bibinfo
  {year} {2012})}\BibitemShut {NoStop}%
\bibitem [{\citenamefont {Paton}\ \emph {et~al.}(2014)\citenamefont {Paton},
  \citenamefont {Varrla}, \citenamefont {Backes}, \citenamefont {Smith},
  \citenamefont {Khan}, \citenamefont {O'Neill}, \citenamefont {Boland},
  \citenamefont {Lotya}, \citenamefont {Istrate}, \citenamefont {King},
  \citenamefont {Higgins}, \citenamefont {Barwich}, \citenamefont {May},
  \citenamefont {Puczkarski}, \citenamefont {Ahmed}, \citenamefont {Moebius},
  \citenamefont {Pettersson}, \citenamefont {Long}, \citenamefont {Coelho},
  \citenamefont {O'Brien}, \citenamefont {McGuire}, \citenamefont {Sanchez},
  \citenamefont {Duesberg}, \citenamefont {McEvoy}, \citenamefont {Pennycook},
  \citenamefont {Downing}, \citenamefont {Crossley}, \citenamefont {Nicolosi},\
  and\ \citenamefont {Coleman}}]{Paton2014}%
  \BibitemOpen
  \bibfield  {author} {\bibinfo {author} {\bibfnamefont {K.~R.}\ \bibnamefont
  {Paton}}, \bibinfo {author} {\bibfnamefont {E.}~\bibnamefont {Varrla}},
  \bibinfo {author} {\bibfnamefont {C.}~\bibnamefont {Backes}}, \bibinfo
  {author} {\bibfnamefont {R.~J.}\ \bibnamefont {Smith}}, \bibinfo {author}
  {\bibfnamefont {U.}~\bibnamefont {Khan}}, \bibinfo {author} {\bibfnamefont
  {A.}~\bibnamefont {O'Neill}}, \bibinfo {author} {\bibfnamefont
  {C.}~\bibnamefont {Boland}}, \bibinfo {author} {\bibfnamefont
  {M.}~\bibnamefont {Lotya}}, \bibinfo {author} {\bibfnamefont {O.~M.}\
  \bibnamefont {Istrate}}, \bibinfo {author} {\bibfnamefont {P.}~\bibnamefont
  {King}}, \bibinfo {author} {\bibfnamefont {T.}~\bibnamefont {Higgins}},
  \bibinfo {author} {\bibfnamefont {S.}~\bibnamefont {Barwich}}, \bibinfo
  {author} {\bibfnamefont {P.}~\bibnamefont {May}}, \bibinfo {author}
  {\bibfnamefont {P.}~\bibnamefont {Puczkarski}}, \bibinfo {author}
  {\bibfnamefont {I.}~\bibnamefont {Ahmed}}, \bibinfo {author} {\bibfnamefont
  {M.}~\bibnamefont {Moebius}}, \bibinfo {author} {\bibfnamefont
  {H.}~\bibnamefont {Pettersson}}, \bibinfo {author} {\bibfnamefont
  {E.}~\bibnamefont {Long}}, \bibinfo {author} {\bibfnamefont {J.}~\bibnamefont
  {Coelho}}, \bibinfo {author} {\bibfnamefont {S.~E.}\ \bibnamefont {O'Brien}},
  \bibinfo {author} {\bibfnamefont {E.~K.}\ \bibnamefont {McGuire}}, \bibinfo
  {author} {\bibfnamefont {B.~M.}\ \bibnamefont {Sanchez}}, \bibinfo {author}
  {\bibfnamefont {G.~S.}\ \bibnamefont {Duesberg}}, \bibinfo {author}
  {\bibfnamefont {N.}~\bibnamefont {McEvoy}}, \bibinfo {author} {\bibfnamefont
  {T.~J.}\ \bibnamefont {Pennycook}}, \bibinfo {author} {\bibfnamefont
  {C.}~\bibnamefont {Downing}}, \bibinfo {author} {\bibfnamefont
  {A.}~\bibnamefont {Crossley}}, \bibinfo {author} {\bibfnamefont
  {V.}~\bibnamefont {Nicolosi}}, \ and\ \bibinfo {author} {\bibfnamefont
  {J.~N.}\ \bibnamefont {Coleman}},\ }\bibfield  {title} {\enquote {\bibinfo
  {title} {{Scalable production of large quantities of defect-free few-layer
  graphene by shear exfoliation in liquids}},}\ }\href {\doibase
  10.1038/nmat3944} {\bibfield  {journal} {\bibinfo  {journal} {Nature
  Materials}\ }\textbf {\bibinfo {volume} {13}},\ \bibinfo {pages} {624--630}
  (\bibinfo {year} {2014})}\BibitemShut {NoStop}%
\bibitem [{\citenamefont {Humphrey}, \citenamefont {Dalke},\ and\ \citenamefont
  {Schulten}(1996)}]{VMD}%
  \BibitemOpen
  \bibfield  {author} {\bibinfo {author} {\bibfnamefont {W.}~\bibnamefont
  {Humphrey}}, \bibinfo {author} {\bibfnamefont {A.}~\bibnamefont {Dalke}}, \
  and\ \bibinfo {author} {\bibfnamefont {K.}~\bibnamefont {Schulten}},\
  }\bibfield  {title} {\enquote {\bibinfo {title} {{VMD -- Visual Molecular
  Dynamics}},}\ }\href {http://www.ks.uiuc.edu/Research/vmd/} {\bibfield
  {journal} {\bibinfo  {journal} {J. Molec. Graphics}\ }\textbf {\bibinfo
  {volume} {14}},\ \bibinfo {pages} {33--38} (\bibinfo {year}
  {1996})}\BibitemShut {NoStop}%
\bibitem [{\citenamefont {Singh}\ \emph
  {et~al.}(2014{\natexlab{a}})\citenamefont {Singh}, \citenamefont {Koch},
  \citenamefont {Subramanian},\ and\ \citenamefont {Stroock}}]{Singh2014}%
  \BibitemOpen
  \bibfield  {author} {\bibinfo {author} {\bibfnamefont {V.}~\bibnamefont
  {Singh}}, \bibinfo {author} {\bibfnamefont {D.~L.}\ \bibnamefont {Koch}},
  \bibinfo {author} {\bibfnamefont {G.}~\bibnamefont {Subramanian}}, \ and\
  \bibinfo {author} {\bibfnamefont {A.~D.}\ \bibnamefont {Stroock}},\
  }\bibfield  {title} {\enquote {\bibinfo {title} {{Rotational motion of a thin
  axisymmetric disk in a low Reynolds number linear flow}},}\ }\href {\doibase
  10.1063/1.4868520} {\bibfield  {journal} {\bibinfo  {journal} {Physics of
  Fluids}\ }\textbf {\bibinfo {volume} {26}} (\bibinfo {year}
  {2014}{\natexlab{a}}),\ 10.1063/1.4868520}\BibitemShut {NoStop}%
\bibitem [{\citenamefont {Ravula}\ \emph {et~al.}(2015)\citenamefont {Ravula},
  \citenamefont {Baker}, \citenamefont {Kamath},\ and\ \citenamefont
  {Baker}}]{Ravula2015}%
  \BibitemOpen
  \bibfield  {author} {\bibinfo {author} {\bibfnamefont {S.}~\bibnamefont
  {Ravula}}, \bibinfo {author} {\bibfnamefont {S.~N.}\ \bibnamefont {Baker}},
  \bibinfo {author} {\bibfnamefont {G.}~\bibnamefont {Kamath}}, \ and\ \bibinfo
  {author} {\bibfnamefont {G.~A.}\ \bibnamefont {Baker}},\ }\bibfield  {title}
  {\enquote {\bibinfo {title} {{Ionic liquid-assisted exfoliation and
  dispersion: Stripping graphene and its two-dimensional layered inorganic
  counterparts of their inhibitions}},}\ }\href {\doibase 10.1039/c4nr01524j}
  {\bibfield  {journal} {\bibinfo  {journal} {Nanoscale}\ }\textbf {\bibinfo
  {volume} {7}},\ \bibinfo {pages} {4338--4353} (\bibinfo {year}
  {2015})}\BibitemShut {NoStop}%
\bibitem [{\citenamefont {Wang}\ \emph {et~al.}(2009)\citenamefont {Wang},
  \citenamefont {Zhang}, \citenamefont {Abidi},\ and\ \citenamefont
  {Cabrales}}]{Wang2009}%
  \BibitemOpen
  \bibfield  {author} {\bibinfo {author} {\bibfnamefont {S.}~\bibnamefont
  {Wang}}, \bibinfo {author} {\bibfnamefont {Y.}~\bibnamefont {Zhang}},
  \bibinfo {author} {\bibfnamefont {N.}~\bibnamefont {Abidi}}, \ and\ \bibinfo
  {author} {\bibfnamefont {L.}~\bibnamefont {Cabrales}},\ }\bibfield  {title}
  {\enquote {\bibinfo {title} {{Wettability and surface free energy of graphene
  films}},}\ }\href {\doibase 10.1021/la901402f} {\bibfield  {journal}
  {\bibinfo  {journal} {Langmuir}\ }\textbf {\bibinfo {volume} {25}},\ \bibinfo
  {pages} {11078--11081} (\bibinfo {year} {2009})}\BibitemShut {NoStop}%
\bibitem [{\citenamefont {van Engers}\ \emph {et~al.}(2017)\citenamefont {van
  Engers}, \citenamefont {Cousens}, \citenamefont {Grobert}, \citenamefont
  {Zappone},\ and\ \citenamefont {Perkin}}]{Babenko2017}%
  \BibitemOpen
  \bibfield  {author} {\bibinfo {author} {\bibfnamefont {C.~D.}\ \bibnamefont
  {van Engers}}, \bibinfo {author} {\bibfnamefont {N.~E.~A.}\ \bibnamefont
  {Cousens}}, \bibinfo {author} {\bibfnamefont {N.}~\bibnamefont {Grobert}},
  \bibinfo {author} {\bibfnamefont {B.}~\bibnamefont {Zappone}}, \ and\
  \bibinfo {author} {\bibfnamefont {S.}~\bibnamefont {Perkin}},\ }\bibfield
  {title} {\enquote {\bibinfo {title} {{Direct Measurement of the Surface
  Energy of Graphene}},}\ }\href {\doibase 10.1021/acs.nanolett.7b01181}
  {\bibfield  {journal} {\bibinfo  {journal} {Nano Letters}\ }\textbf {\bibinfo
  {volume} {17}},\ \bibinfo {pages} {3815--3821} (\bibinfo {year}
  {2017})}\BibitemShut {NoStop}%
\bibitem [{\citenamefont {Hernandez}\ \emph {et~al.}(2010)\citenamefont
  {Hernandez}, \citenamefont {Lotya}, \citenamefont {Rickard}, \citenamefont
  {Bergin},\ and\ \citenamefont {Coleman}}]{Hernandez2010}%
  \BibitemOpen
  \bibfield  {author} {\bibinfo {author} {\bibfnamefont {Y.}~\bibnamefont
  {Hernandez}}, \bibinfo {author} {\bibfnamefont {M.}~\bibnamefont {Lotya}},
  \bibinfo {author} {\bibfnamefont {D.}~\bibnamefont {Rickard}}, \bibinfo
  {author} {\bibfnamefont {S.~D.}\ \bibnamefont {Bergin}}, \ and\ \bibinfo
  {author} {\bibfnamefont {J.~N.}\ \bibnamefont {Coleman}},\ }\bibfield
  {title} {\enquote {\bibinfo {title} {{Measurement of multicomponent
  solubility parameters for graphene facilitates solvent discovery}},}\ }\href
  {\doibase 10.1021/la903188a} {\bibfield  {journal} {\bibinfo  {journal}
  {Langmuir}\ }\textbf {\bibinfo {volume} {26}},\ \bibinfo {pages} {3208--3213}
  (\bibinfo {year} {2010})}\BibitemShut {NoStop}%
\bibitem [{\citenamefont {An}\ \emph {et~al.}(2010)\citenamefont {An},
  \citenamefont {Simmons}, \citenamefont {Shah}, \citenamefont {Wolfe},
  \citenamefont {Lewis}, \citenamefont {Washington}, \citenamefont {Nayak},
  \citenamefont {Talapatra},\ and\ \citenamefont {Kar}}]{An2010}%
  \BibitemOpen
  \bibfield  {author} {\bibinfo {author} {\bibfnamefont {X.}~\bibnamefont
  {An}}, \bibinfo {author} {\bibfnamefont {T.}~\bibnamefont {Simmons}},
  \bibinfo {author} {\bibfnamefont {R.}~\bibnamefont {Shah}}, \bibinfo {author}
  {\bibfnamefont {C.}~\bibnamefont {Wolfe}}, \bibinfo {author} {\bibfnamefont
  {K.~M.}\ \bibnamefont {Lewis}}, \bibinfo {author} {\bibfnamefont
  {M.}~\bibnamefont {Washington}}, \bibinfo {author} {\bibfnamefont {S.~K.}\
  \bibnamefont {Nayak}}, \bibinfo {author} {\bibfnamefont {S.}~\bibnamefont
  {Talapatra}}, \ and\ \bibinfo {author} {\bibfnamefont {S.}~\bibnamefont
  {Kar}},\ }\bibfield  {title} {\enquote {\bibinfo {title} {{Stable aqueous
  dispersions of noncovalently functionalized graphene from graphite and their
  multifunctional high-performance applications}},}\ }\href {\doibase
  10.1021/nl903557p} {\bibfield  {journal} {\bibinfo  {journal} {Nano Letters}\
  }\textbf {\bibinfo {volume} {10}},\ \bibinfo {pages} {4295--4301} (\bibinfo
  {year} {2010})}\BibitemShut {NoStop}%
\bibitem [{\citenamefont {Shih}\ \emph {et~al.}(2010)\citenamefont {Shih},
  \citenamefont {Shangchao}, \citenamefont {Strano},\ and\ \citenamefont
  {Blankschtein}}]{Shih2010}%
  \BibitemOpen
  \bibfield  {author} {\bibinfo {author} {\bibfnamefont {C.~J.}\ \bibnamefont
  {Shih}}, \bibinfo {author} {\bibfnamefont {L.}~\bibnamefont {Shangchao}},
  \bibinfo {author} {\bibfnamefont {M.}~\bibnamefont {Strano}}, \ and\ \bibinfo
  {author} {\bibfnamefont {D.}~\bibnamefont {Blankschtein}},\ }\bibfield
  {title} {\enquote {\bibinfo {title} {{Understanding the Stabilization of
  Liquid-Phase-Exfoliated Graphene in Polar Solvents: Molecular Dynamics
  Simulations and Kinetic Theory of Colloid Aggregation Chih-Jen}},}\ }\href
  {\doibase 10.1021/nl903557p} {\bibfield  {journal} {\bibinfo  {journal} {Nano
  Letters}\ }\textbf {\bibinfo {volume} {10}},\ \bibinfo {pages} {4295--4301}
  (\bibinfo {year} {2010})}\BibitemShut {NoStop}%
\bibitem [{\citenamefont {Sresht}, \citenamefont {P{\'{a}}dua},\ and\
  \citenamefont {Blankschtein}(2015)}]{Sresht2015}%
  \BibitemOpen
  \bibfield  {author} {\bibinfo {author} {\bibfnamefont {V.}~\bibnamefont
  {Sresht}}, \bibinfo {author} {\bibfnamefont {A.~A.}\ \bibnamefont
  {P{\'{a}}dua}}, \ and\ \bibinfo {author} {\bibfnamefont {D.}~\bibnamefont
  {Blankschtein}},\ }\bibfield  {title} {\enquote {\bibinfo {title}
  {{Liquid-Phase Exfoliation of Phosphorene: Design Rules from Molecular
  Dynamics Simulations}},}\ }\href {\doibase 10.1021/acsnano.5b02683}
  {\bibfield  {journal} {\bibinfo  {journal} {ACS Nano}\ }\textbf {\bibinfo
  {volume} {9}},\ \bibinfo {pages} {8255--8268} (\bibinfo {year}
  {2015})}\BibitemShut {NoStop}%
\bibitem [{\citenamefont {Bordes}, \citenamefont {Szala-Bilnik},\ and\
  \citenamefont {Padua}(2018)}]{Bordes2018a}%
  \BibitemOpen
  \bibfield  {author} {\bibinfo {author} {\bibfnamefont {E.}~\bibnamefont
  {Bordes}}, \bibinfo {author} {\bibfnamefont {J.}~\bibnamefont
  {Szala-Bilnik}}, \ and\ \bibinfo {author} {\bibfnamefont {A.~H.}\
  \bibnamefont {Padua}},\ }\bibfield  {title} {\enquote {\bibinfo {title}
  {{Exfoliation of graphene and fluorographene in molecular and ionic
  liquids}},}\ }\href {\doibase 10.1039/c7fd00169j} {\bibfield  {journal}
  {\bibinfo  {journal} {Faraday Discussions}\ }\textbf {\bibinfo {volume}
  {206}},\ \bibinfo {pages} {61--75} (\bibinfo {year} {2018})}\BibitemShut
  {NoStop}%
\bibitem [{\citenamefont {Plimpton}(1995)}]{LAMMPS}%
  \BibitemOpen
  \bibfield  {author} {\bibinfo {author} {\bibfnamefont {S.}~\bibnamefont
  {Plimpton}},\ }\bibfield  {title} {\enquote {\bibinfo {title} {{Fast Parallel
  Algorithms For Short-range Molecular-dynamics}},}\ }\href {\doibase
  doi:10.1006/jcph.1995.1039} {\bibfield  {journal} {\bibinfo  {journal} {J.
  Comp. Phys.}\ }\textbf {\bibinfo {volume} {117}},\ \bibinfo {pages} {1--19}
  (\bibinfo {year} {1995})}\BibitemShut {NoStop}%
\bibitem [{\citenamefont {Stuart}\ \emph {et~al.}(2000)\citenamefont {Stuart},
  \citenamefont {Tutein}, \citenamefont {Harrison},\ and\ \citenamefont
  {Introduction}}]{Stuart2000}%
  \BibitemOpen
  \bibfield  {author} {\bibinfo {author} {\bibfnamefont {S.~J.}\ \bibnamefont
  {Stuart}}, \bibinfo {author} {\bibfnamefont {A.~B.}\ \bibnamefont {Tutein}},
  \bibinfo {author} {\bibfnamefont {J.~A.}\ \bibnamefont {Harrison}}, \ and\
  \bibinfo {author} {\bibfnamefont {I.}~\bibnamefont {Introduction}},\
  }\bibfield  {title} {\enquote {\bibinfo {title} {{A reactive potential for
  hydrocarbons with intermolecular interactions}},}\ }\href@noop {} {\bibfield
  {journal} {\bibinfo  {journal} {Journal of Chemical Physics}\ }\textbf
  {\bibinfo {volume} {112}},\ \bibinfo {pages} {6472--6486} (\bibinfo {year}
  {2000})}\BibitemShut {NoStop}%
\bibitem [{\citenamefont {Abascal}\ and\ \citenamefont {Vega}(2005)}]{TIP4P}%
  \BibitemOpen
  \bibfield  {author} {\bibinfo {author} {\bibfnamefont {J.~L.}\ \bibnamefont
  {Abascal}}\ and\ \bibinfo {author} {\bibfnamefont {C.}~\bibnamefont {Vega}},\
  }\bibfield  {title} {\enquote {\bibinfo {title} {{A general purpose model for
  the condensed phases of water: TIP4P/2005.}}}\ }\href {\doibase
  10.1063/1.2121687} {\bibfield  {journal} {\bibinfo  {journal} {The Journal of
  chemical physics}\ }\textbf {\bibinfo {volume} {123}},\ \bibinfo {pages}
  {234505} (\bibinfo {year} {2005})}\BibitemShut {NoStop}%
\bibitem [{\citenamefont {Schmid}\ \emph {et~al.}(2011)\citenamefont {Schmid},
  \citenamefont {Eichenberger}, \citenamefont {Choutko}, \citenamefont
  {Riniker}, \citenamefont {Winger}, \citenamefont {Mark},\ and\ \citenamefont
  {{Van Gunsteren}}}]{Schmid2011}%
  \BibitemOpen
  \bibfield  {author} {\bibinfo {author} {\bibfnamefont {N.}~\bibnamefont
  {Schmid}}, \bibinfo {author} {\bibfnamefont {A.~P.}\ \bibnamefont
  {Eichenberger}}, \bibinfo {author} {\bibfnamefont {A.}~\bibnamefont
  {Choutko}}, \bibinfo {author} {\bibfnamefont {S.}~\bibnamefont {Riniker}},
  \bibinfo {author} {\bibfnamefont {M.}~\bibnamefont {Winger}}, \bibinfo
  {author} {\bibfnamefont {A.~E.}\ \bibnamefont {Mark}}, \ and\ \bibinfo
  {author} {\bibfnamefont {W.~F.}\ \bibnamefont {{Van Gunsteren}}},\ }\bibfield
   {title} {\enquote {\bibinfo {title} {{Definition and testing of the GROMOS
  force-field versions 54A7 and 54B7}},}\ }\href {\doibase
  10.1007/s00249-011-0700-9} {\bibfield  {journal} {\bibinfo  {journal}
  {European Biophysics Journal}\ }\textbf {\bibinfo {volume} {40}},\ \bibinfo
  {pages} {843--856} (\bibinfo {year} {2011})}\BibitemShut {NoStop}%
\bibitem [{\citenamefont {Malde}\ \emph {et~al.}(2011)\citenamefont {Malde},
  \citenamefont {Zuo}, \citenamefont {Breeze}, \citenamefont {Stroet},
  \citenamefont {Poger}, \citenamefont {Nair}, \citenamefont {Oostenbrink},\
  and\ \citenamefont {Mark}}]{Malde2011}%
  \BibitemOpen
  \bibfield  {author} {\bibinfo {author} {\bibfnamefont {A.~K.}\ \bibnamefont
  {Malde}}, \bibinfo {author} {\bibfnamefont {L.}~\bibnamefont {Zuo}}, \bibinfo
  {author} {\bibfnamefont {M.}~\bibnamefont {Breeze}}, \bibinfo {author}
  {\bibfnamefont {M.}~\bibnamefont {Stroet}}, \bibinfo {author} {\bibfnamefont
  {D.}~\bibnamefont {Poger}}, \bibinfo {author} {\bibfnamefont {P.~C.}\
  \bibnamefont {Nair}}, \bibinfo {author} {\bibfnamefont {C.}~\bibnamefont
  {Oostenbrink}}, \ and\ \bibinfo {author} {\bibfnamefont {A.~E.}\ \bibnamefont
  {Mark}},\ }\bibfield  {title} {\enquote {\bibinfo {title} {{An Automated
  force field Topology Builder (ATB) and repository: Version 1.0}},}\ }\href
  {\doibase 10.1021/ct200196m} {\bibfield  {journal} {\bibinfo  {journal}
  {Journal of Chemical Theory and Computation}\ }\textbf {\bibinfo {volume}
  {7}},\ \bibinfo {pages} {4026--4037} (\bibinfo {year} {2011})}\BibitemShut
  {NoStop}%
\bibitem [{\citenamefont {Nose}(1984)}]{Nose1984}%
  \BibitemOpen
  \bibfield  {author} {\bibinfo {author} {\bibfnamefont {S.}~\bibnamefont
  {Nose}},\ }\bibfield  {title} {\enquote {\bibinfo {title} {{A molecular
  dynamics method for simulations in the canonical ensemble}},}\ }\href
  {\doibase 10.1080/00268978400101201} {\bibfield  {journal} {\bibinfo
  {journal} {Mol. Phys.}\ }\textbf {\bibinfo {volume} {52}},\ \bibinfo {pages}
  {255--268} (\bibinfo {year} {1984})}\BibitemShut {NoStop}%
\bibitem [{\citenamefont {Hoover}(1985)}]{Hoover1985}%
  \BibitemOpen
  \bibfield  {author} {\bibinfo {author} {\bibfnamefont {W.~G.}\ \bibnamefont
  {Hoover}},\ }\bibfield  {title} {\enquote {\bibinfo {title} {{Canonical
  dynamics: equilibrium phase-space distributions}},}\ }\href {\doibase
  10.1103/PhysRevA.31.1695} {\bibfield  {journal} {\bibinfo  {journal}
  {Physical Review A}\ }\textbf {\bibinfo {volume} {31}},\ \bibinfo {pages}
  {1695--1697} (\bibinfo {year} {1985})}\BibitemShut {NoStop}%
\bibitem [{\citenamefont {Gonz{\'{a}}lez}\ and\ \citenamefont
  {Abascal}(2010)}]{Gonzalez2010}%
  \BibitemOpen
  \bibfield  {author} {\bibinfo {author} {\bibfnamefont {M.~A.}\ \bibnamefont
  {Gonz{\'{a}}lez}}\ and\ \bibinfo {author} {\bibfnamefont {J.~L.~F.}\
  \bibnamefont {Abascal}},\ }\bibfield  {title} {\enquote {\bibinfo {title}
  {{The shear viscosity of rigid water models}},}\ }\href {\doibase
  10.1063/1.3330544} {\bibfield  {journal} {\bibinfo  {journal} {Journal of
  Chemical Physics}\ }\textbf {\bibinfo {volume} {132}},\ \bibinfo {pages}
  {096101} (\bibinfo {year} {2010})}\BibitemShut {NoStop}%
\bibitem [{\citenamefont {Henni}\ \emph {et~al.}(2004)\citenamefont {Henni},
  \citenamefont {Hromek}, \citenamefont {Tontiwachwuthikul},\ and\
  \citenamefont {Chakma}}]{Henni2004}%
  \BibitemOpen
  \bibfield  {author} {\bibinfo {author} {\bibfnamefont {A.}~\bibnamefont
  {Henni}}, \bibinfo {author} {\bibfnamefont {J.~J.}\ \bibnamefont {Hromek}},
  \bibinfo {author} {\bibfnamefont {P.}~\bibnamefont {Tontiwachwuthikul}}, \
  and\ \bibinfo {author} {\bibfnamefont {A.}~\bibnamefont {Chakma}},\
  }\bibfield  {title} {\enquote {\bibinfo {title} {{Volumetric properties and
  viscosities for aqueous N-Methyl-2-pyrrolidone solutions from 25°C to
  70°C}},}\ }\href {\doibase 10.1021/je034073k} {\bibfield  {journal}
  {\bibinfo  {journal} {Journal of Chemical and Engineering Data}\ }\textbf
  {\bibinfo {volume} {49}},\ \bibinfo {pages} {231--234} (\bibinfo {year}
  {2004})}\BibitemShut {NoStop}%
\bibitem [{\citenamefont {Henry}\ \emph {et~al.}(2005)\citenamefont {Henry},
  \citenamefont {Lukey}, \citenamefont {Evans},\ and\ \citenamefont
  {Yarovsky}}]{Henry2005}%
  \BibitemOpen
  \bibfield  {author} {\bibinfo {author} {\bibfnamefont {D.~J.}\ \bibnamefont
  {Henry}}, \bibinfo {author} {\bibfnamefont {C.~A.}\ \bibnamefont {Lukey}},
  \bibinfo {author} {\bibfnamefont {E.}~\bibnamefont {Evans}}, \ and\ \bibinfo
  {author} {\bibfnamefont {I.}~\bibnamefont {Yarovsky}},\ }\bibfield  {title}
  {\enquote {\bibinfo {title} {{Theoretical study of adhesion between graphite,
  polyester and silica surfaces}},}\ }\href {\doibase
  10.1080/089270412331332712} {\bibfield  {journal} {\bibinfo  {journal}
  {Molecular Simulation}\ }\textbf {\bibinfo {volume} {31}},\ \bibinfo {pages}
  {449--455} (\bibinfo {year} {2005})}\BibitemShut {NoStop}%
\bibitem [{\citenamefont {Ferguson}\ and\ \citenamefont
  {Irons}(1941)}]{Ferguson1941}%
  \BibitemOpen
  \bibfield  {author} {\bibinfo {author} {\bibfnamefont {A.}~\bibnamefont
  {Ferguson}}\ and\ \bibinfo {author} {\bibfnamefont {E.~J.}\ \bibnamefont
  {Irons}},\ }\bibfield  {title} {\enquote {\bibinfo {title} {{On surface
  energy and surface entropy}},}\ }\href {\doibase 10.1088/0959-5309/53/2/309}
  {\bibfield  {journal} {\bibinfo  {journal} {Proceedings of the Physical
  Society}\ }\textbf {\bibinfo {volume} {53}},\ \bibinfo {pages} {182--185}
  (\bibinfo {year} {1941})}\BibitemShut {NoStop}%
\bibitem [{\citenamefont {L{\'{o}}pez}\ \emph {et~al.}(2013)\citenamefont
  {L{\'{o}}pez}, \citenamefont {Garc{\'{i}}a-Abu{\'{i}}n}, \citenamefont
  {G{\'{o}}mez-D{\'{i}}az}, \citenamefont {{La Rubia}},\ and\ \citenamefont
  {Navaza}}]{Lopez2013}%
  \BibitemOpen
  \bibfield  {author} {\bibinfo {author} {\bibfnamefont {A.~B.}\ \bibnamefont
  {L{\'{o}}pez}}, \bibinfo {author} {\bibfnamefont {A.}~\bibnamefont
  {Garc{\'{i}}a-Abu{\'{i}}n}}, \bibinfo {author} {\bibfnamefont
  {D.}~\bibnamefont {G{\'{o}}mez-D{\'{i}}az}}, \bibinfo {author} {\bibfnamefont
  {M.~D.}\ \bibnamefont {{La Rubia}}}, \ and\ \bibinfo {author} {\bibfnamefont
  {J.~M.}\ \bibnamefont {Navaza}},\ }\bibfield  {title} {\enquote {\bibinfo
  {title} {{Density, speed of sound, viscosity, refractive index and surface
  tension of N-methyl-2-pyrrolidone + diethanolamine (or triethanolamine) from
  T = (293.15 to 323.15) K}},}\ }\href {\doibase 10.1016/j.jct.2013.01.020}
  {\bibfield  {journal} {\bibinfo  {journal} {Journal of Chemical
  Thermodynamics}\ }\textbf {\bibinfo {volume} {61}},\ \bibinfo {pages} {1--6}
  (\bibinfo {year} {2013})}\BibitemShut {NoStop}%
\bibitem [{\citenamefont {Alejandre}\ and\ \citenamefont
  {Chapela}(2010)}]{Alejandre2010}%
  \BibitemOpen
  \bibfield  {author} {\bibinfo {author} {\bibfnamefont {J.}~\bibnamefont
  {Alejandre}}\ and\ \bibinfo {author} {\bibfnamefont {G.~A.}\ \bibnamefont
  {Chapela}},\ }\bibfield  {title} {\enquote {\bibinfo {title} {{The surface
  tension of TIP4P/2005 water model using the Ewald sums for the dispersion
  interactions}},}\ }\href {\doibase 10.1063/1.3279128} {\bibfield  {journal}
  {\bibinfo  {journal} {Journal of Chemical Physics}\ }\textbf {\bibinfo
  {volume} {132}} (\bibinfo {year} {2010}),\ 10.1063/1.3279128}\BibitemShut
  {NoStop}%
\bibitem [{\citenamefont {Kannam}\ \emph {et~al.}(2013)\citenamefont {Kannam},
  \citenamefont {Todd}, \citenamefont {Hansen},\ and\ \citenamefont
  {Daivis}}]{Kannam2013}%
  \BibitemOpen
  \bibfield  {author} {\bibinfo {author} {\bibfnamefont {S.~K.}\ \bibnamefont
  {Kannam}}, \bibinfo {author} {\bibfnamefont {B.~D.}\ \bibnamefont {Todd}},
  \bibinfo {author} {\bibfnamefont {J.~S.}\ \bibnamefont {Hansen}}, \ and\
  \bibinfo {author} {\bibfnamefont {P.~J.}\ \bibnamefont {Daivis}},\ }\bibfield
   {title} {\enquote {\bibinfo {title} {{How fast does water flow in carbon
  nanotubes ?}}}\ }\href@noop {} {\bibfield  {journal} {\bibinfo  {journal}
  {The Journal of chemical physics}\ }\textbf {\bibinfo {volume} {094701}},\
  \bibinfo {pages} {1--9} (\bibinfo {year} {2013})}\BibitemShut {NoStop}%
\bibitem [{\citenamefont {Lauga}, \citenamefont {Brenner},\ and\ \citenamefont
  {Stone}(2007)}]{Lauga2007}%
  \BibitemOpen
  \bibfield  {author} {\bibinfo {author} {\bibfnamefont {E.}~\bibnamefont
  {Lauga}}, \bibinfo {author} {\bibfnamefont {M.}~\bibnamefont {Brenner}}, \
  and\ \bibinfo {author} {\bibfnamefont {H.}~\bibnamefont {Stone}},\ }\bibfield
   {title} {\enquote {\bibinfo {title} {{Microfluidics: The No-Slip Boundary
  Condition}},}\ }\href {\doibase 10.1007/978-3-540-30299-5_19} {\bibfield
  {journal} {\bibinfo  {journal} {Springer Handbook of Experimental Fluid
  Mechanics}\ ,\ \bibinfo {pages} {1219--1240}} (\bibinfo {year}
  {2007})}\BibitemShut {NoStop}%
\bibitem [{\citenamefont {Bocquet}\ and\ \citenamefont
  {Barrat}(2006)}]{Bocquet2007}%
  \BibitemOpen
  \bibfield  {author} {\bibinfo {author} {\bibfnamefont {L.}~\bibnamefont
  {Bocquet}}\ and\ \bibinfo {author} {\bibfnamefont {J.~L.}\ \bibnamefont
  {Barrat}},\ }\bibfield  {title} {\enquote {\bibinfo {title} {{Flow boundary
  conditions from nano- to micro-scales}},}\ }\href {\doibase 10.1039/b616490k}
  {\bibfield  {journal} {\bibinfo  {journal} {Soft Matter}\ }\textbf {\bibinfo
  {volume} {3}},\ \bibinfo {pages} {685--693} (\bibinfo {year} {2006})},\
  \Eprint {http://arxiv.org/abs/0612242} {arXiv:0612242 [cond-mat]}
  \BibitemShut {NoStop}%
\bibitem [{\citenamefont {Pozrikidis}(1992)}]{pozrikidis1992boundary}%
  \BibitemOpen
  \bibfield  {author} {\bibinfo {author} {\bibfnamefont {C.}~\bibnamefont
  {Pozrikidis}},\ }\bibfield  {title} {\enquote {\bibinfo {title} {{Boundary
  Integral and Singularity Methods for Linearized Viscous Flow, Cambridge
  University Press, Cambridge, 1992}},}\ }\href@noop {} {\  (\bibinfo {year}
  {1992})}\BibitemShut {NoStop}%
\bibitem [{\citenamefont {Kamal}, \citenamefont {Gravelle},\ and\ \citenamefont
  {Botto}()}]{Kamal2019}%
  \BibitemOpen
  \bibfield  {author} {\bibinfo {author} {\bibfnamefont {C.}~\bibnamefont
  {Kamal}}, \bibinfo {author} {\bibfnamefont {S.}~\bibnamefont {Gravelle}}, \
  and\ \bibinfo {author} {\bibfnamefont {L.}~\bibnamefont {Botto}},\ }\bibfield
   {title} {\enquote {\bibinfo {title} {{Graphene nanoplatelets attain a stable
  orientation in a shear flow}},}\ }\href@noop {} {\bibinfo  {journal} {under
  consideration}\ }\BibitemShut {NoStop}%
\bibitem [{\citenamefont {Maali}, \citenamefont {Cohen-bouhacina},\ and\
  \citenamefont {Kellay}(2008)}]{Maali2008}%
  \BibitemOpen
\bibfield  {journal} {  }\bibfield  {author} {\bibinfo {author} {\bibfnamefont
  {A.}~\bibnamefont {Maali}}, \bibinfo {author} {\bibfnamefont
  {T.}~\bibnamefont {Cohen-bouhacina}}, \ and\ \bibinfo {author} {\bibfnamefont
  {H.}~\bibnamefont {Kellay}},\ }\bibfield  {title} {\enquote {\bibinfo {title}
  {{Measurement of the slip length of water flow on graphite surface}},}\
  }\href {\doibase 10.1063/1.2840717} {\bibfield  {journal} {\bibinfo
  {journal} {Applied Physics Letters}\ }\textbf {\bibinfo {volume} {92}}
  (\bibinfo {year} {2008}),\ 10.1063/1.2840717}\BibitemShut {NoStop}%
\bibitem [{\citenamefont {Ortiz-Young}\ \emph {et~al.}(2013)\citenamefont
  {Ortiz-Young}, \citenamefont {Chiu}, \citenamefont {Kim}, \citenamefont
  {Vo{\"{i}}tchovsky},\ and\ \citenamefont {Riedo}}]{Ortiz-Young2013}%
  \BibitemOpen
  \bibfield  {author} {\bibinfo {author} {\bibfnamefont {D.}~\bibnamefont
  {Ortiz-Young}}, \bibinfo {author} {\bibfnamefont {H.-C.}\ \bibnamefont
  {Chiu}}, \bibinfo {author} {\bibfnamefont {S.}~\bibnamefont {Kim}}, \bibinfo
  {author} {\bibfnamefont {K.}~\bibnamefont {Vo{\"{i}}tchovsky}}, \ and\
  \bibinfo {author} {\bibfnamefont {E.}~\bibnamefont {Riedo}},\ }\bibfield
  {title} {\enquote {\bibinfo {title} {{The interplay between apparent
  viscosity and wettability in nanoconfined water}},}\ }\href {\doibase
  10.1038/ncomms3482} {\bibfield  {journal} {\bibinfo  {journal} {Nature
  communications}\ } (\bibinfo {year} {2013}),\ 10.1038/ncomms3482}\BibitemShut
  {NoStop}%
\bibitem [{\citenamefont {Tocci}, \citenamefont {Joly},\ and\ \citenamefont
  {Michaelides}(2014)}]{Tocci2014}%
  \BibitemOpen
  \bibfield  {author} {\bibinfo {author} {\bibfnamefont {G.}~\bibnamefont
  {Tocci}}, \bibinfo {author} {\bibfnamefont {L.}~\bibnamefont {Joly}}, \ and\
  \bibinfo {author} {\bibfnamefont {A.}~\bibnamefont {Michaelides}},\
  }\bibfield  {title} {\enquote {\bibinfo {title} {{Friction of water on
  graphene and hexagonal BN from ab initio methods : very different slippage
  despite very similar interface structures}},}\ }\href {\doibase
  10.1021/nl502837d} {\bibfield  {journal} {\bibinfo  {journal} {Nano letters}\
  }\textbf {\bibinfo {volume} {14}},\ \bibinfo {pages} {6872--6877} (\bibinfo
  {year} {2014})}\BibitemShut {NoStop}%
\bibitem [{\citenamefont {Singh}\ \emph
  {et~al.}(2014{\natexlab{b}})\citenamefont {Singh}, \citenamefont {Koch},
  \citenamefont {Subramanian},\ and\ \citenamefont {Stroock}}]{Singh2014a}%
  \BibitemOpen
  \bibfield  {author} {\bibinfo {author} {\bibfnamefont {V.}~\bibnamefont
  {Singh}}, \bibinfo {author} {\bibfnamefont {D.~L.}\ \bibnamefont {Koch}},
  \bibinfo {author} {\bibfnamefont {G.}~\bibnamefont {Subramanian}}, \ and\
  \bibinfo {author} {\bibfnamefont {A.~D.}\ \bibnamefont {Stroock}},\
  }\bibfield  {title} {\enquote {\bibinfo {title} {{Rotational motion of a thin
  axisymmetric disk in a low Reynolds number linear flow}},}\ }\href {\doibase
  10.1063/1.4868520} {\bibfield  {journal} {\bibinfo  {journal} {Physics of
  Fluids}\ }\textbf {\bibinfo {volume} {26}} (\bibinfo {year}
  {2014}{\natexlab{b}}),\ 10.1063/1.4868520}\BibitemShut {NoStop}%
\bibitem [{\citenamefont {Shen}\ \emph {et~al.}(2015)\citenamefont {Shen},
  \citenamefont {He}, \citenamefont {Wu}, \citenamefont {Gao}, \citenamefont
  {Keyshar}, \citenamefont {Zhang}, \citenamefont {Yang}, \citenamefont {Ye},
  \citenamefont {Vajtai}, \citenamefont {Lou},\ and\ \citenamefont
  {Ajayan}}]{Shen2015}%
  \BibitemOpen
  \bibfield  {author} {\bibinfo {author} {\bibfnamefont {J.}~\bibnamefont
  {Shen}}, \bibinfo {author} {\bibfnamefont {Y.}~\bibnamefont {He}}, \bibinfo
  {author} {\bibfnamefont {J.}~\bibnamefont {Wu}}, \bibinfo {author}
  {\bibfnamefont {C.}~\bibnamefont {Gao}}, \bibinfo {author} {\bibfnamefont
  {K.}~\bibnamefont {Keyshar}}, \bibinfo {author} {\bibfnamefont
  {X.}~\bibnamefont {Zhang}}, \bibinfo {author} {\bibfnamefont
  {Y.}~\bibnamefont {Yang}}, \bibinfo {author} {\bibfnamefont {M.}~\bibnamefont
  {Ye}}, \bibinfo {author} {\bibfnamefont {R.}~\bibnamefont {Vajtai}}, \bibinfo
  {author} {\bibfnamefont {J.}~\bibnamefont {Lou}}, \ and\ \bibinfo {author}
  {\bibfnamefont {P.~M.}\ \bibnamefont {Ajayan}},\ }\bibfield  {title}
  {\enquote {\bibinfo {title} {{Liquid Phase Exfoliation of Two-Dimensional
  Materials by Directly Probing and Matching Surface Tension Components}},}\
  }\href {\doibase 10.1021/acs.nanolett.5b01842} {\bibfield  {journal}
  {\bibinfo  {journal} {Nano Letters}\ }\textbf {\bibinfo {volume} {15}},\
  \bibinfo {pages} {5449--5454} (\bibinfo {year} {2015})}\BibitemShut {NoStop}%
\bibitem [{\citenamefont {Kirkwood}\ and\ \citenamefont
  {Buff}(1949)}]{Kirkwood1949}%
  \BibitemOpen
  \bibfield  {author} {\bibinfo {author} {\bibfnamefont {J.~G.}\ \bibnamefont
  {Kirkwood}}\ and\ \bibinfo {author} {\bibfnamefont {F.~P.}\ \bibnamefont
  {Buff}},\ }\bibfield  {title} {\enquote {\bibinfo {title} {{The statistical
  mechanical theory of surface tension}},}\ }\href {\doibase 10.1063/1.1747248}
  {\bibfield  {journal} {\bibinfo  {journal} {The Journal of Chemical Physics}\
  }\textbf {\bibinfo {volume} {17}},\ \bibinfo {pages} {338--343} (\bibinfo
  {year} {1949})}\BibitemShut {NoStop}%
\bibitem [{\citenamefont {Dreher}\ \emph {et~al.}(2018)\citenamefont {Dreher},
  \citenamefont {Lemarchand}, \citenamefont {Soulard}, \citenamefont
  {Bourasseau}, \citenamefont {Malfreyt},\ and\ \citenamefont
  {Pineau}}]{Dreher2018a}%
  \BibitemOpen
  \bibfield  {author} {\bibinfo {author} {\bibfnamefont {T.}~\bibnamefont
  {Dreher}}, \bibinfo {author} {\bibfnamefont {C.}~\bibnamefont {Lemarchand}},
  \bibinfo {author} {\bibfnamefont {L.}~\bibnamefont {Soulard}}, \bibinfo
  {author} {\bibfnamefont {E.}~\bibnamefont {Bourasseau}}, \bibinfo {author}
  {\bibfnamefont {P.}~\bibnamefont {Malfreyt}}, \ and\ \bibinfo {author}
  {\bibfnamefont {N.}~\bibnamefont {Pineau}},\ }\bibfield  {title} {\enquote
  {\bibinfo {title} {{Calculation of a solid/liquid surface tension: A
  methodological study}},}\ }\href {\doibase 10.1063/1.5008473} {\bibfield
  {journal} {\bibinfo  {journal} {Journal of Chemical Physics}\ }\textbf
  {\bibinfo {volume} {148}} (\bibinfo {year} {2018}),\
  10.1063/1.5008473}\BibitemShut {NoStop}%
\bibitem [{\citenamefont {Dreher}\ \emph {et~al.}(2019)\citenamefont {Dreher},
  \citenamefont {Lemarchand}, \citenamefont {Pineau}, \citenamefont
  {Bourasseau}, \citenamefont {Ghoufi},\ and\ \citenamefont
  {Malfreyt}}]{Dreher2019b}%
  \BibitemOpen
  \bibfield  {author} {\bibinfo {author} {\bibfnamefont {T.}~\bibnamefont
  {Dreher}}, \bibinfo {author} {\bibfnamefont {C.}~\bibnamefont {Lemarchand}},
  \bibinfo {author} {\bibfnamefont {N.}~\bibnamefont {Pineau}}, \bibinfo
  {author} {\bibfnamefont {E.}~\bibnamefont {Bourasseau}}, \bibinfo {author}
  {\bibfnamefont {A.}~\bibnamefont {Ghoufi}}, \ and\ \bibinfo {author}
  {\bibfnamefont {P.}~\bibnamefont {Malfreyt}},\ }\bibfield  {title} {\enquote
  {\bibinfo {title} {{Calculation of the interfacial tension of the
  graphene-water interaction by molecular simulations}},}\ }\href {\doibase
  10.1063/1.5048576} {\bibfield  {journal} {\bibinfo  {journal} {Journal of
  Chemical Physics}\ }\textbf {\bibinfo {volume} {150}} (\bibinfo {year}
  {2019}),\ 10.1063/1.5048576}\BibitemShut {NoStop}%
\bibitem [{\citenamefont {Karagiannidis}\ \emph {et~al.}(2017)\citenamefont
  {Karagiannidis}, \citenamefont {Hodge}, \citenamefont {Lombardi},
  \citenamefont {Tomarchio}, \citenamefont {Decorde}, \citenamefont {Milana},
  \citenamefont {Goykhman}, \citenamefont {Su}, \citenamefont {Mesite},
  \citenamefont {Johnstone}, \citenamefont {Leary}, \citenamefont {Midgley},
  \citenamefont {Pugno}, \citenamefont {Torrisi},\ and\ \citenamefont
  {Ferrari}}]{Karagiannidis2017}%
  \BibitemOpen
  \bibfield  {author} {\bibinfo {author} {\bibfnamefont {P.~G.}\ \bibnamefont
  {Karagiannidis}}, \bibinfo {author} {\bibfnamefont {S.~A.}\ \bibnamefont
  {Hodge}}, \bibinfo {author} {\bibfnamefont {L.}~\bibnamefont {Lombardi}},
  \bibinfo {author} {\bibfnamefont {F.}~\bibnamefont {Tomarchio}}, \bibinfo
  {author} {\bibfnamefont {N.}~\bibnamefont {Decorde}}, \bibinfo {author}
  {\bibfnamefont {S.}~\bibnamefont {Milana}}, \bibinfo {author} {\bibfnamefont
  {I.}~\bibnamefont {Goykhman}}, \bibinfo {author} {\bibfnamefont
  {Y.}~\bibnamefont {Su}}, \bibinfo {author} {\bibfnamefont {S.~V.}\
  \bibnamefont {Mesite}}, \bibinfo {author} {\bibfnamefont {D.~N.}\
  \bibnamefont {Johnstone}}, \bibinfo {author} {\bibfnamefont {R.~K.}\
  \bibnamefont {Leary}}, \bibinfo {author} {\bibfnamefont {P.~A.}\ \bibnamefont
  {Midgley}}, \bibinfo {author} {\bibfnamefont {N.~M.}\ \bibnamefont {Pugno}},
  \bibinfo {author} {\bibfnamefont {F.}~\bibnamefont {Torrisi}}, \ and\
  \bibinfo {author} {\bibfnamefont {A.~C.}\ \bibnamefont {Ferrari}},\
  }\bibfield  {title} {\enquote {\bibinfo {title} {{Microfluidization of
  Graphite and Formulation of Graphene-Based Conductive Inks}},}\ }\href
  {\doibase 10.1021/acsnano.6b07735} {\bibfield  {journal} {\bibinfo  {journal}
  {ACS Nano}\ }\textbf {\bibinfo {volume} {11}},\ \bibinfo {pages} {2742--2755}
  (\bibinfo {year} {2017})}\BibitemShut {NoStop}%
\bibitem [{\citenamefont {Lingard}\ and\ \citenamefont
  {Whitmore}(1974)}]{Lingard1974}%
  \BibitemOpen
  \bibfield  {author} {\bibinfo {author} {\bibfnamefont {P.~S.}\ \bibnamefont
  {Lingard}}\ and\ \bibinfo {author} {\bibfnamefont {R.~L.}\ \bibnamefont
  {Whitmore}},\ }\bibfield  {title} {\enquote {\bibinfo {title} {{The
  deformation of disc-shaped particles by a shearing fluid with application to
  the red blood cell}},}\ }\href {\doibase 10.1016/0021-9797(74)90306-3}
  {\bibfield  {journal} {\bibinfo  {journal} {Journal of Colloid And Interface
  Science}\ }\textbf {\bibinfo {volume} {49}},\ \bibinfo {pages} {119--127}
  (\bibinfo {year} {1974})}\BibitemShut {NoStop}%
\bibitem [{\citenamefont {Botto}(2019)}]{botto2019towards}%
  \BibitemOpen
  \bibfield  {author} {\bibinfo {author} {\bibfnamefont {L.}~\bibnamefont
  {Botto}},\ }\bibfield  {title} {\enquote {\bibinfo {title} {{Towards
  nanomechanical models of liquid-phase exfoliation of layered 2D
  nanomaterials: analysis of a pi-peel model}},}\ }\href@noop {} {\bibfield
  {journal} {\bibinfo  {journal} {Frontiers in Materials}\ }\textbf {\bibinfo
  {volume} {6}},\ \bibinfo {pages} {302} (\bibinfo {year} {2019})}\BibitemShut
  {NoStop}%
\bibitem [{\citenamefont {Salussolia}\ \emph {et~al.}(2019)\citenamefont
  {Salussolia}, \citenamefont {Pugno}, \citenamefont {Barbieri},\ and\
  \citenamefont {Botto}}]{Salussolia2019}%
  \BibitemOpen
  \bibfield  {author} {\bibinfo {author} {\bibfnamefont {G.}~\bibnamefont
  {Salussolia}}, \bibinfo {author} {\bibfnamefont {N.}~\bibnamefont {Pugno}},
  \bibinfo {author} {\bibfnamefont {E.}~\bibnamefont {Barbieri}}, \ and\
  \bibinfo {author} {\bibfnamefont {L.}~\bibnamefont {Botto}},\ }\bibfield
  {title} {\enquote {\bibinfo {title} {{Micromechanics of liquid-phase
  exfoliation of a layered 2D material: a hydrodynamic peeling model}},}\
  }\href {\doibase 10.1016/j.jmps.2019.103764} {\bibfield  {journal} {\bibinfo
  {journal} {J. Mech. Phys. Solids}\ } (\bibinfo {year} {2019}),\
  10.1016/j.jmps.2019.103764}\BibitemShut {NoStop}%
\end{thebibliography}%

\end{document}